\documentclass[12pt]{article}
\begin{document}
\title{Finite viscoelasticity of filled rubbers:
the effects of pre-loading and thermal recovery}

\author{Aleksey D. Drozdov and Al Dorfmann\\
Institute of Structural Engineering\\
82 Peter Jordan Street\\
1190 Vienna, Austria}
\date{}
\maketitle

\begin{abstract}
Constitutive equations are derived for the viscoelastic behavior
of filled elastomers at isothermal loading with finite strains. A
particle-reinforced rubber is thought of as a composite where
regions with low concentrations of junctions between chains are
randomly distributed in the bulk material. The onset of these
inclusions is associated with the inhomogeneity in spatial
distribution of a cross-linker during the mixing process. With
reference to the theory of transient networks, the time-dependent
response of an elastomer is modelled as thermally activated
processes of breakage and reformation of chains in domains with
low concentrations of junctions, whereas junctions in the bulk
medium are treated as permanent. Stress--strain relations are
developed by using the laws of thermodynamics. Adjustable
parameters in the constitutive equations are found by fitting
experimental data in tensile relaxation tests for several grades
of unfilled and carbon black filled natural rubber. It is
demonstrated that (i) the average relaxation time noticeably grows
with the elongation ratio, which is explained by
mechanically-induced crystallization of strands, and (ii) the
relaxation spectrum of a filled elastomer is not affected by
mechanical pre-loading and thermal recovery at elevated
temperatures.
\end{abstract}
\newpage

\section{Introduction}

This paper is concerned with the viscoelastic behavior of unfilled
and particle-reinforced rubbers at isothermal loading with finite
strains. The time-dependent response of filled elastomers has
attracted substantial attention in the past decade
\cite{AH96,BB98,CET00,Gon93,GS92,HS98,HS01,HS95,HS96,JS93,JQF95,LRK93,Lio96,Lio97,Lio98,MK00,OB92,RG98,SE98,Spa97,WL00},
which is explained by numerous industrial applications of these
materials (vehicle tires, shock absorbers, earthquake bearings,
seals, flexible joints, solid propellants, etc.).

Observations in uniaxial and biaxial tests evidence that rubbers
reinforced with carbon black or silica, as well as solid
propellants filled with hard particles reveal strong
time-dependent response
\cite{BB98,HS01,HS95,Lio96,Lio97,Lio98,MK00,SE98,WL00}.

A particle-reinforced rubber is conventionally thought of as a
network of macromolecules bridged by junctions (chemical and
physical cross-links, entanglements and filler particles). The
classical theory of rubber elasticity \cite{Tre75} treats all
junctions as permanent, which implies that chains cannot detach
from the junctions. To predict the time-dependent response of
rubbery polymers, Green and Tobolsky \cite{GT46} presumed that
chains can slip from the junctions as they are thermally agitated.
The concept of temporary junctions was developed in
\cite{Lod68,TE92,Yam56} and was applied to the description of the
viscoelastic behavior of rubbers in \cite{Dro97,ES99}.

We treat an elastomer as a strongly inhomogeneous material and
assume that the characteristic length of an inhomogeneity
substantially exceeds the radius of gyration for a macromolecule,
on the one hand, and it is noticeably less than a size of a
specimen, on the other.

To describe the viscoelastic response of a rubber (unfilled or
particle-reinforced), we model it as a composite medium, where
regions with low concentrations of junctions (RLCJ) are randomly
distributed in the bulk material. These domains arise during the
preparation of rubber samples (at the stages of mixing and
vulcanization) because of the spatial inhomogeneity in the
distribution of a cross-linker. We treat junctions in the host
matrix as permanent and associate the time-dependent response of
elastomers with breakage and reformation of chains in inclusions
with low concentrations of junctions. The qualitative difference
in the behavior of chains in RLCJs and in the bulk medium is
explained by the fact that surrounding macromolecules in the bulk
material prevent slippage of chains from the junctions, whereas
they do not affect the slippage process in RLCJs, where
interactions between chains are noticeably weaker.

According to the theory of temporary networks \cite{TE92}, a
strand whose ends are linked to separate neighboring junctions is
treated as an active one. Snapping of an end of a strand from a
junction is thought of as its breakage (transition from the active
state of the strand to its dangling state). When the free end of a
dangling strand captures a junction (which reflects creation of a
new physical cross-link or entanglement), a new active strand
merges with the network. Breakage and reformation of active
strands are treated as thermally activated processes: attachment
and detachment events occur at random times as the strands are
agitated by thermal fluctuations.

The objective of this paper is three-fold:
\begin{enumerate}
\item
to derive constitutive equations for the time-dependent behavior
of particle-reinforc\-ed rubbers using the concept of temporary
networks,

\item to carry out uniaxial tensile relaxation tests on
virgin specimens of unfilled and particle-reinforced natural
rubber, as well as on the same specimens after mechanical
pre-loading and recovery at an elevated temperature,

\item to apply the stress--strain relations to the
analysis of experimental data.
\end{enumerate}
Our purpose is to evaluate the effects of pre-loading and thermal
recovery on the average rate of stress relaxation and on the
relaxation spectrum.

The study is organized as follows. Breakage and reformation of
active strands in the regions with low concentrations of junctions
are discussed in Section 2. Governing equations for the kinetics
of rearrangement are introduced in Section 3. The strain energy
density is determined in Section 4 for the bulk material, in
Section 5 for RLCJs and in Section 6 for the composite medium.
Stress--strain relations at finite strains are derived in Section
7 by using the laws of thermodynamics. Uniaxial extension of a
specimen is analyzed in Section 8. Section 9 deals with the
description of tensile relaxation tests. Adjustable parameters in
the constitutive equations are found in Section 10 by fitting
observations. A brief discussion of results is presented in
Section 11. Some concluding remarks are formulated in Section 12.

\section{Breakage and reformation of active strands}

Breakage and reformation of strands in RLCJs are modelled as
thermally activated processes. Slippage of an end of an active
strand from a junction and reformation of a dangling strand occur
at random times when they are thermally agitated. With reference
to the theory of thermally activated processes \cite{Eyr36}, the
rate of breakage for an active strand in a stress-free medium is
given by
\begin{equation}
\Gamma=\Gamma_{\ast} \exp \Bigl (-\frac{\tilde{\omega}}{k_{\rm
B}T} \Bigr ),
\end{equation}
where $\tilde{\omega}>0$ is the energy for breakage, $k_{\rm B}$
is Boltzmann's constant, and the pre-factor $\Gamma_{\ast}$ weakly
depends on temperature. Equation (1) implies that for any
$\tilde{\omega}$, the rate of breakage vanishes at low
temperatures ($T\to 0$) and it approaches $\Gamma_{\ast}$ at
elevated temperatures ($T\to\infty$).

The dimensionless energy for breakage reads
\[ \omega=\frac{\tilde{\omega}}{k_{\rm B}T^{\circ}}, \]
where $T^{\circ}$ is some reference temperature. Disregarding the
effect of small increments of temperature, $\Delta T=T-T^{\circ}$,
on the rate of breakage, $\Gamma$, we find the rate of transition
from the active state of a strand to its dangling state
\[ \Gamma=\Gamma_{\ast} \exp (-\omega).\]
In what follows, this equation is assumed to be satisfied for an
arbitrary loading program, provided that the coefficient
$\Gamma_{\ast}$ depends on current strains and some internal
variables that characterize
\begin{enumerate}
\item
damage of the secondary network and isolated clusters of filler,

\item
changes in the structure of macromolecules driven by their
alignment and mechanic\-al\-ly-induced crystallization.
\end{enumerate}
This implies that for a time-dependent loading, the rate of
breakage for active strands reads
\begin{equation}
\Gamma(t,\omega)=\Gamma_{\ast}(t) \exp (-\omega),
\end{equation}
where the function $\Gamma_{\ast}(t)$ is found by fitting
observations.

Denote by $p(\omega)$ the distribution function for energies of
breakage for active strands in RLCJs. The function $p(\omega)$ is
assumed to be robust with respect to the influence of
thermo-mechanical factors. This means that $p$ is a function which
does not depend on temperature (for small increments, $\Delta T$)
and on the history of loading. To fit experimental data, we apply
the quasi-Gaussian distribution
\begin{equation}
p(\omega) = p_{0}\exp \biggl
[-\frac{(\omega-\Omega)^{2}}{2\Sigma^{2}}\biggr ],
\end{equation}
where $\Omega$ and $\Sigma$ are adjustable parameters, and the
constant $p_{0}$ is found from the condition
\begin{equation}
\int_{0}^{\infty} p(\omega) d\omega =1.
\end{equation}
As it will be demonstrated later, the ratio
\[ \xi=\frac{\Sigma}{\Omega}, \]
that characterizes the width of the quasi-Gaussian distribution
(3), is relatively small (about 0.5), which implies that $\Omega$
may be associated with the average energy for breakage of an
active strand, and $\Sigma$ determines the standard deviation for
energies of breakage.

The function $\Gamma_{\ast}(t)$ in Eq. (2) and the parameters
$\Omega$ and $\Sigma$ in Eq. (3) entirely describe slippage of
active strands from temporary junctions in the regions with low
concentrations of cross-links.

\section{Kinetic equations for rearrangement of strands}

The kinetics of breakage and reformation of active strands is
determined by the function $X(t,\tau,\omega)$, which equals the
current (at time $t$) number (per unit mass) of active strands in
RLCJs  with the energy for breakage $\omega$ that have last been
bridged to the network before an instant $\tau\in [0,t]$ (the
initial time $t=0$ corresponds to the instant when external loads
are applied to the specimen).

The quantity $X(t,t,\omega)$ equals the number of active strands
(per unit mass) with the energy for breakage $\omega$ at time $t$.
In particular, $X(0,0,\omega)$ is the initial number of active
strands with the energy for breakage $\omega$,
\begin{equation}
X(0,0,\omega)=N_{\rm L} p(\omega),
\end{equation}
where $N_{\rm L}$ is the average number of active strands per unit
mass of RLCJs. The amount
\[ \frac{\partial X}{\partial \tau}(t,\tau,\omega) \biggl |_{t=\tau} \; d\tau \]
is the number (per unit mass) of active strands with the energy
for breakage $\omega$ that have been linked to the network within
the interval $[\tau,\tau+d\tau ]$. The quantity
\[ \frac{\partial X}{\partial \tau} (t,\tau,\omega)\;d\tau \]
determines the number of these strands that have not been broken
during the interval $[\tau, t]$. The amount
\[ -\frac{\partial X}{\partial t} (t,0,\omega)\;dt \]
is the number of active strands (per unit mass) that detach from
the network (for the first time) within the interval $[t,t+dt ]$,
and the quantity
\[ -\frac{\partial^{2} X}{\partial t\partial \tau} (t,\tau,\omega)\;dtd\tau \]
is the number of strands (per unit mass) that have last been
linked to the network within the interval $[\tau,\tau+d\tau ]$ and
slip from temporary junctions (for the first time after merging)
during the interval $[t,t+dt ]$.

The evolution for a transient network is determined by the
relative rate of breakage for active strands, $\Gamma(t,\omega)$,
and by the rate of merging with the network for dangling strands,
$\gamma(t,\omega)$. The assumption that $\Gamma$ and $\gamma$
depend on the energy for breakage, $\omega$, distinguishes the
present model from previous ones. The dependence of the rates of
breakage and reformation on time, $t$, reflects the effects of
strains and mechanically-induced changes in the internal structure
of chains on these parameters under a time-dependent loading.

The rate of breakage, $\Gamma$, equals the ratio of the number of
active strands broken per unit time to the total number of active
strands. It follows from this definition that for strands
connected with the network before loading and remaining active at
time $t$,
\begin{equation}
\Gamma(t,\omega) = -\frac{1}{X(t,0,\omega)}\frac{\partial
X}{\partial t}(t,0,\omega).
\end{equation}
Applying the definition of the rate of breakage to strands merged
with the network at time $\tau\geq 0$ and remaining active at time
$t\geq \tau$, we find that
\begin{equation}
\Gamma(t,\omega) = -\biggl [ \frac{\partial X}{\partial
\tau}(t,\tau,\omega) \biggr ]^{-1} \frac{\partial^{2} X}{\partial
t\partial \tau}(t,\tau,\omega).
\end{equation}
It is assumed in Eqs. (6) and (7) that the rate of breakage,
$\Gamma$, is independent of the instant, $\tau$, when an active
strand merges with the network.

The rate of reformation, $\gamma$, equals the number of dangling
strands (per unit mass) bridged to the network per unit time
\begin{equation}
\gamma(t,\omega)=\frac{\partial X}{\partial
\tau}(t,\tau,\omega)\biggl |_{\tau=t}.
\end{equation}
Integration of differential equations (6) and (7) with initial
conditions (5) and (8) implies that
\begin{eqnarray}
X(t,0,\omega) &=& N_{\rm L} p(\omega)\exp \biggl [ -\int_{0}^{t}
\Gamma(s,\omega) ds\biggr ],
\nonumber\\
\frac{\partial X}{\partial \tau}(t,\tau,\omega) &=&
\gamma(\tau,\omega)\exp \biggl [-\int_{\tau}^{t} \Gamma(s,\omega)
ds \biggr ].
\end{eqnarray}
To exclude the rate of reformation, $\gamma$, from Eq. (9), we
suppose that the concentrations of active and dangling strands are
independent of time. This means that the number of strands broken
per unit time coincides with the number of strands merging with
the network within the same interval. The number of active strands
detached from the network during the interval $[t,t+dt ]$ equals
the sum of the number of initial strands (not broken until time
$t$) that slip from temporary junctions
\[ -\frac{\partial X}{\partial t}(t,0,\omega)\; dt \]
and the number of strands linked with the network within the
interval $[\tau, \tau+d\tau ]$ (for the last time before instant
$t$) for various instants $\tau$ and broken within the interval
$[t,t+dt ]$
\[ -\frac{\partial^{2} X}{\partial t\partial \tau}(t,\tau,\omega)\;dt d\tau. \]
This number coincides with the number of strands,
$\gamma(t,\omega)\; dt$, merged with the network within the
interval $[t,t+dt]$, which results in the balance law
\[
\gamma(t,\omega)=-\frac{\partial X}{\partial t}(t,0,\omega)
-\int_{0}^{t} \frac{\partial^{2} X}{\partial t\partial \tau}
(t,\tau,\omega) d\tau .
\]
Substitution of Eq. (9) into this equality implies that
\begin{equation}
\gamma(t,\omega) = \Gamma (t,\omega)\biggl \{ N_{\rm L}
p(\omega)\exp \biggl [-\int_{0}^{t} \Gamma(s,\omega) ds \biggr ]
+\int_{0}^{t} \gamma(\tau,\omega) \exp \biggl [-\int_{\tau}^{t}
\Gamma(s,\omega) ds \biggr ] d\tau\biggr \}.
\end{equation}
The solution of the linear integral equation (10) reads
\begin{equation}
\gamma(t,\omega)=N_{\rm L} p(\omega)\Gamma(t,\omega).
\end{equation}
Equations (9) and (11) determine the current distribution of
active strands in the regions with low concentrations of
junctions.

\section{Mechanical energy of the bulk material}

The bulk medium is treated as a permanent network of long chains
bridged by junctions. We adopt the affinity hypothesis which
disregards thermal oscillations of junctions and presumes that the
deformation gradient for the motion of junctions at the
micro-level coincides with the deformation gradient for the motion
of appropriate points of a sample at the macro-level.

The bulk material is modelled as an isotropic incompressible
medium, whose strain energy density, $W_{\rm B}$, depends on the
current temperature $T$ (the strong effect of temperature reflects
the entropic nature of the mechanical energy for elastomers), a
vector of internal variables ${\bf a}$ (which accounts for changes
in the structure of a particle-reinforced rubber driven by rupture
of filler clusters, stress-induced alignment of macromolecules,
and their crystallization) and the first two principal invariants
$I_{k}$ ($k=1,2$) of the Cauchy deformation tensor
\[
{\bf C}^{\circ}(t)={\bf \nabla}^{\circ} {\bf R}(t)\cdot \Bigl
[{\bf \nabla}^{\circ} {\bf R}(t)\Bigr ]^{\top},
\]
where ${\bf \nabla}^{\circ} {\bf R}(t)$ is the deformation
gradient at time $t$, ${\bf \nabla}^{\circ}$ is the gradient
operator in the reference state, the dot stands for inner product,
and $\top$ denotes transpose. The strain energy density reads
\begin{equation}
W_{\rm B}(t)=\bar{W}_{\rm B}\Bigl (T(t), {\bf a}(t),I_{1}({\bf
C}^{\circ}(t)), I_{2}({\bf C}^{\circ}(t))\Bigr ),
\end{equation}
where the function $\bar{W}_{\rm B}$ obeys the condition
\begin{equation}
\bar{W}_{\rm B}(T,{\bf a},I_{1},I_{2})\Bigl |_{I_{1}=3,
I_{2}=3}=0.
\end{equation}
Equation (13) means that the mechanical energy vanishes in the
stress-free state. In this equation we disregard thermally induced
changes in the reference state of rubber compared to the strains
under consideration.

\section{Mechanical energy of RLCJs}

Regions with low concentration of junctions are assumed to be
randomly distributed in the bulk material as inclusions with
various sizes and shapes. Any RLCJ is modelled as an isotropic and
incompressible transient network, where breakage and reformation
of active strands are driven by thermal fluctuations. The strain
energy of an active strand in regions with low concentration of
junctions, $w_{\rm L}$, is assumed to depend on temperature, $T$,
the vector of internal variables, ${\bf a}$, and the first two
principal invariants of the Cauchy deformation tensor. The
conventional hypothesis that stress in a dangling strand totally
relaxes before this strand captures a new junction implies that
the stress-free state of a strand merging with the network at time
$\tau\geq 0$ coincides with the deformed state of the network at
that instant. This means that for a strand not broken within the
interval $[0,t]$,
\[
w_{\rm L}(t,0)=\bar{w}_{\rm L}\Bigl (T(t), {\bf a}(t),I_{1}({\bf
C}^{\circ}(t)), I_{2}({\bf C}^{\circ}(t))\Bigr ),
\]
and for a strand that has last been reformed at time $\tau\in
[0,t]$,
\[
w_{\rm L}(t,\tau)=\bar{w}_{\rm L}\Bigl (T(t), {\bf
a}(t),I_{1}({\bf C}(t,\tau)),I_{2}({\bf C}(t,\tau))\Bigr ),
\]
where ${\bf C}(t,\tau)$ is the relative Cauchy deformation tensor for transition
from the deformed state of a specimen at instant $\tau$ (when the strand
has last been merged to the network) to its deformed state at time $t\geq \tau$.
The tensor ${\bf C}(t,\tau)$ is expressed in terms of the deformation gradient
${\bf \nabla}(\tau){\bf R}(t)$ by the formula
\[
{\bf C}(t,\tau)={\bf \nabla}(\tau) {\bf R}(t)\cdot \Bigl [{\bf
\nabla}(\tau) {\bf R}(t)\Bigr ]^{\top},
\]
where ${\bf \nabla}(\tau)$ is the gradient operator in the
deformed state of the medium at time $\tau$.

Summing mechanical energies for active strands with various
energies for breakage, $\omega$, merged with the network at
various instants, $\tau\in [0,t]$, and neglecting the energy of
interaction between strands, we arrive at the strain energy
density per unit mass of RLCJs,
\[
W_{\rm L}(t)=\int_{0}^{\infty} \biggl [ X(t,0,\omega)w_{\rm L}(t,0)
+\int_{0}^{t} \frac{\partial X}{\partial \tau}(t,\tau,\omega)w_{\rm L}(t,\tau)
d\tau \biggr ] d\omega.
\]
Substitution of expressions (9) and (11) into this equality
implies that
\begin{eqnarray}
W_{\rm L}(t) &=& \int_{0}^{\infty} p(\omega)d\omega
\biggl \{ \bar{W}_{\rm L}\Bigl (T(t), I_{1}({\bf C}^{\circ}(t)),
I_{2}({\bf C}^{\circ}(t))\Bigr )\exp \Bigl [-\int_{0}^{t} \Gamma(s,\omega) ds\Bigr ]
\nonumber\\
&&+\int_{0}^{t} \Gamma(\tau,\omega)
\bar{W}_{\rm L}\Bigl (T(t), I_{1}({\bf C}(t,\tau)),I_{2}({\bf C}(t,\tau))\Bigr )
\nonumber\\
&&\times
\exp \Bigl [-\int_{\tau}^{t} \Gamma(s,\omega) ds\Bigr ] d\tau \biggr \},
\end{eqnarray}
where the function $\bar{W}_{\rm L}(T, I_{1},I_{2})=N_{\rm
L}\bar{w}_{\rm L}(T, I_{1},I_{2})$ satisfies condition
\begin{equation}
\bar{W}_{\rm L}(T,{\bf a},I_{1},I_{2})\Bigl |_{I_{1}=3, I_{2}=3}=0
\end{equation}
similar to Eq. (13). Formula (14) determines the strain energy
density (per unit mass) of inclusions with low concentrations of
junctions.

\section{Strain energy of a filled rubber}

Denote by $\mu_{\rm L}$ the mass fraction of regions with low
concentrations of junctions. Taking into account that the strain
energy density of a composite material, $W$, is an extensive
thermodynamic variable, we find that
\begin{equation}
W=(1-\mu_{\rm L})W_{\rm B}+\mu_{\rm L}W_{\rm L}.
\end{equation}
Substitution of expressions (12) and (14) into Eq. (16) results in
the formula
\begin{eqnarray}
W(t)&=& (1-\mu_{\rm L})\bar{W}_{\rm B}\Bigl (T(t), {\bf a}(t),
I_{1}({\bf C}^{\circ}(t)), I_{2}({\bf C}^{\circ}(t))\Bigr )
\nonumber\\
&&+\mu_{\rm L} \int_{0}^{\infty} p(\omega)d\omega \biggl \{
\bar{W}_{\rm L}\Bigl (T(t), {\bf a}(t), I_{1}({\bf C}^{\circ}(t)),
I_{2}({\bf C}^{\circ}(t))\Bigr )\exp \Bigl [-\int_{0}^{t}
\Gamma(s,\omega) ds\Bigr ]
\nonumber\\
&&+\int_{0}^{t} \Gamma(\tau,\omega) \bar{W}_{\rm L}\Bigl (T(t),
{\bf a}(t), I_{1}({\bf C}(t,\tau)),I_{2}({\bf C}(t,\tau))\Bigr )
\nonumber\\
&&\times \exp \Bigl [-\int_{\tau}^{t} \Gamma(s,\omega) ds\Bigr ]
d\tau \biggr \}.
\end{eqnarray}
It follows from Eqs. (15) and (17) that
\begin{equation}
\frac{dW}{dt}(t)=J(t)\frac{dT}{dt}(t)+{\bf A}(t)\cdot\frac{d{\bf
a}}{dt}(t) +U_{1}^{\circ}(t)\frac{dI_{1}}{dt}({\bf C}^{\circ}(t))
+U_{2}^{\circ}(t)\frac{dI_{2}}{dt}({\bf C}^{\circ}(t)) +U(t)-Y(t),
\end{equation}
where
\begin{eqnarray}
J(t)&=& (1-\mu_{\rm L})\frac{\partial \bar{W}_{\rm B}}{\partial T}
\Bigl (T(t), {\bf a}(t), I_{1}({\bf C}^{\circ}(t)), I_{2}({\bf
C}^{\circ}(t))\Bigr )
\nonumber\\
&&+\mu_{\rm L} \int_{0}^{\infty} p(\omega)d\omega \biggl \{
\frac{\partial \bar{W}_{\rm L}}{\partial T} \Bigl (T(t), {\bf
a}(t), I_{1}({\bf C}^{\circ}(t)),I_{2}({\bf C}^{\circ}(t))\Bigr )
\exp \Bigl [-\int_{0}^{t} \Gamma(s,\omega) ds\Bigr ]
\nonumber\\
&&+\int_{0}^{t} \Gamma(\tau,\omega)\frac{\partial \bar{W}_{\rm
L}}{\partial T} \Bigl (T(t), {\bf a}(t), I_{1}({\bf
C}(t,\tau)),I_{2}({\bf C}(t,\tau))\Bigr ) \exp \Bigl
[-\int_{\tau}^{t} \Gamma(s,\omega) ds\Bigr ] d\tau \biggr \},
\nonumber\\
{\bf A}(t) &=& (1-\mu_{\rm L})\frac{\partial \bar{W}_{\rm
B}}{\partial {\bf a}} \Bigl (T(t), {\bf a}(t), I_{1}({\bf
C}^{\circ}(t)), I_{2}({\bf C}^{\circ}(t))\Bigr )
\nonumber\\
&&+\mu_{\rm L} \int_{0}^{\infty} p(\omega)d\omega \biggl \{
\frac{\partial \bar{W}_{\rm L}}{\partial {\bf a}} \Bigl (T(t),
{\bf a}(t), I_{1}({\bf C}^{\circ}(t)),I_{2}({\bf
C}^{\circ}(t))\Bigr ) \exp \Bigl [-\int_{0}^{t} \Gamma(s,\omega)
ds\Bigr ]
\nonumber\\
&&+\int_{0}^{t} \Gamma(\tau,\omega)\frac{\partial \bar{W}_{\rm
L}}{\partial {\bf a}} \Bigl (T(t), {\bf a}(t), I_{1}({\bf
C}(t,\tau)),I_{2}({\bf C}(t,\tau))\Bigr ) \exp \Bigl
[-\int_{\tau}^{t} \Gamma(s,\omega) ds\Bigr ] d\tau \biggr \},
\nonumber\\
U_{k}^{\circ}(t)&=& (1-\mu_{\rm L})\frac{\partial \bar{W}_{\rm
B}}{\partial I_{k}} \Bigl (T(t),  {\bf a}(t), I_{1}({\bf
C}^{\circ}(t)), I_{2}({\bf C}^{\circ}(t))\Bigr )
\nonumber\\
&&+\mu_{\rm L} \int_{0}^{\infty} \frac{\partial \bar{W}_{\rm
L}}{\partial I_{k}} \Bigl (T(t), {\bf a}(t), I_{1}({\bf
C}^{\circ}(t)),I_{2}({\bf C}^{\circ}(t))\Bigr ) \exp \Bigl
[-\int_{0}^{t} \Gamma(s,\omega) ds\Bigr ]p(\omega)d\omega,
\nonumber\\
U(t) &=& \mu_{\rm L} \int_{0}^{\infty} p(\omega)d\omega
\int_{0}^{t} \Gamma(\tau,\omega)\exp \Bigl [-\int_{\tau}^{t}
\Gamma(s,\omega) ds\Bigr ] d\tau
\nonumber\\
&&\times
\biggl [ \frac{\partial \bar{W}_{\rm L}}{\partial I_{1}}
\Bigl (T(t),  {\bf a}(t), I_{1}({\bf C}(t,\tau)),I_{2}({\bf
C}(t,\tau))\Bigr ) \frac{\partial I_{1}}{\partial t}({\bf
C}(t,\tau))
\nonumber\\
&& +\frac{\partial \bar{W}_{\rm L}}{\partial I_{2}} \Bigl (T(t),
{\bf a}(t), I_{1}({\bf C}(t,\tau)),I_{2}({\bf C}(t,\tau))\Bigr )
\frac{\partial I_{2}}{\partial t}({\bf C}(t,\tau))\biggr ],
\nonumber\\
Y(t) &=& \mu_{\rm L} \int_{0}^{\infty} \Gamma(t,\omega) p(\omega)
\biggl \{ \bar{W}_{\rm L}\Bigl (T(t), {\bf a}(t), I_{1}({\bf
C}^{\circ}(t)), I_{2}({\bf C}^{\circ}(t))\Bigr )\exp \Bigl
[-\int_{0}^{t} \Gamma(s,\omega) ds\Bigr ]
\nonumber\\
&&+ \int_{0}^{t} \Gamma(\tau,\omega) \bar{W}_{\rm L}\Bigl (T(t),
{\bf a}(t), I_{1}({\bf C}(t,\tau)),I_{2}({\bf C}(t,\tau))\Bigr )
\nonumber\\
&&\times
\exp \Bigl [-\int_{\tau}^{t} \Gamma(s,\omega) ds\Bigr ] d\tau \biggr \}d\omega.
\end{eqnarray}
The principal invariants of the Cauchy deformation tensor, ${\bf C}(t,\tau)$,
coincide with the principal invariants of the Finger tensor for transition
from the deformed state at time $\tau$ to the deformed state at time $t$,
\[ {\bf F}(t,\tau)=\Bigl [{\bf \nabla}(\tau){\bf R}(t)\Bigr ]^{\top}
\cdot {\bf \nabla}(\tau){\bf R}(t). \]
This assertion implies that
\begin{equation}
\frac{\partial I_{k}}{\partial t}({\bf C}(t,\tau))
=\frac{\partial I_{k}}{\partial t}({\bf F}(t,\tau))
=\frac{\partial I_{k}}{\partial {\bf F}}({\bf F}(t,\tau)):
\biggl [ \frac{\partial {\bf F}}{\partial t}(t,\tau)\biggr ]^{\top},
\end{equation}
where the colon stands for convolution of tensors. The derivatives
of the principal invariants, $I_{k}$, read \cite{Dro96}
\[ \frac{\partial I_{1}}{\partial {\bf F}}({\bf F})={\bf I},
\qquad
\frac{\partial I_{2}}{\partial {\bf F}}({\bf F})
=I_{1}({\bf F}){\bf I}-{\bf F}^{\top}, \]
where ${\bf I}$ is the unit tensor.
The derivative of the tensor ${\bf F}$ with respect to time is given by \cite{Dro96}
\begin{equation}
\frac{\partial {\bf F}}{\partial t}(t,\tau) =\Bigl [ {\bf
\nabla}(t){\bf V}(t)\Bigr ]^{\top} \cdot {\bf F}(t,\tau) +{\bf
F}(t,\tau)\cdot{\bf \nabla}(t){\bf V}(t),
\end{equation}
where ${\bf V}$ is the velocity vector. It follows from Eqs. (20)
and (21) that
\begin{eqnarray}
\frac{\partial I_{1}}{\partial t}({\bf C}(t,\tau))
&=& 2{\bf F}(t,\tau):{\bf D}(t),
\nonumber\\
\frac{\partial I_{2}}{\partial t}({\bf C}(t,\tau))
&=& 2\Bigl [ I_{1}({\bf C}(t,\tau)){\bf F}(t,\tau)
-{\bf F}^{2}(t,\tau)\Bigr ]:{\bf D}(t),
\end{eqnarray}
where ${\bf D}(t)=\frac{1}{2}\Bigl [ {\bf \nabla}(t) {\bf V}(t)+
\Bigl ({\bf \nabla}(t){\bf V}(t)\Bigr )^{\top}\Bigr ]$ is the
rate-of-strain tensor. Substitution of expressions (22) into
formula (19) for the functional $U$ implies that
\begin{equation}
U(t)= 2\mu_{\rm L}{\bf \Lambda}_{\rm L}(t):{\bf D}(t).
\end{equation}
Here
\begin{eqnarray*}
{\bf \Lambda}_{\rm L}(t) &=& \int_{0}^{\infty} p(\omega)d\omega
\int_{0}^{t} \Bigl [ \phi_{\rm L\,1}(t,\tau){\bf F}(t,\tau)
+\phi_{\rm L\,2}(t,\tau){\bf F}^{2}(t,\tau)\Bigr ]
\nonumber\\
&&\times \Gamma(\tau,\omega)
\exp \Bigl [-\int_{\tau}^{t} \Gamma(s,\omega) ds\Bigr ] d\tau,
\end{eqnarray*}
and the functions $\phi_{{\rm L}\,k}(t,\tau)$ read
\begin{eqnarray*}
\phi_{\rm L\,1}(t,\tau)&=& \frac{\partial \bar{W}_{\rm L}
}{\partial I_{1}} \Bigl (T(t), {\bf a}(t), I_{1}({\bf
C}(t,\tau)),I_{2}({\bf C}(t,\tau))\Bigr )
\nonumber\\
&& +I_{1}({\bf C}(t,\tau))\frac{\partial \bar{W}_{\rm L}}{\partial
I_{2}} \Bigl (T(t), {\bf a}(t), I_{1}({\bf C}(t,\tau)),I_{2}({\bf
C}(t,\tau))\Bigr ),
\nonumber\\
\phi_{\rm L\,2}(t,\tau) &=& -\frac{\partial \bar{W}_{\rm
L}}{\partial I_{2}} \Bigl (T(t), {\bf a}(t), I_{1}({\bf
C}(t,\tau)),I_{2}({\bf C}(t,\tau))\Bigr ).
\end{eqnarray*}
By analogy with Eq. (23), we find that
\begin{eqnarray}
\sum_{k=1}^{2} U_{k}^{\circ}(t)\frac{dI_{k}}{dt}({\bf
C}^{\circ}(t)) &=& 2\Bigl \{ (1-\mu_{\rm L}){\bf \Lambda}_{\rm
B}^{\circ}(t)
\nonumber\\
&&+\mu_{\rm L}{\bf \Lambda}_{\rm L}^{\circ}(t)\int_{0}^{\infty}
\exp \Bigl [-\int_{0}^{t} \Gamma(s,\omega)ds\Bigr ]
p(\omega)d\omega \Bigr \}:{\bf D}(t),
\end{eqnarray}
where
\[
{\bf \Lambda}_{\rm \beta}^{\circ}(t) =
\phi_{\beta\,1}^{\circ}(t){\bf F}^{\circ}(t)
+\phi_{\beta\,2}^{\circ}(t)\Bigl [ {\bf F}^{\circ}(t)\Bigr ]^{2}
\qquad (\beta={\rm B,L}),
\]
${\bf F}^{\circ}(t)=\Bigl [{\bf \nabla}^{\circ} {\bf R}(t)\Bigr
]^{\top} \cdot {\bf \nabla}^{\circ} {\bf R}(t)$ is the Finger
tensor for transition from the reference state of the medium to
its deformed state at time $t$, and the functions
$\phi_{\beta\,k}^{\circ}(t)$ are given by
\begin{eqnarray*}
\phi_{\beta\,1}^{\circ}(t)&=& \frac{\partial
\bar{W}_{\beta}}{\partial I_{1}} \Bigl (T(t), {\bf a}(t),
I_{1}({\bf C}^{\circ}(t)),I_{2}({\bf C}^{\circ}(t))\Bigr )
\nonumber\\
&& +I_{1}({\bf C}^{\circ}(t)) \frac{\partial
\bar{W}_{\beta}}{\partial I_{2}} \Bigl (T(t), {\bf a}(t),
I_{1}({\bf C}^{\circ}(t)),I_{2}({\bf C}^{\circ}(t))\Bigr ),
\nonumber\\
\phi_{\beta\,2}^{\circ}(t)&=& - \frac{\partial \bar{W}_{\beta
}}{\partial I_{2}} \Bigl (T(t), {\bf a}(t), I_{1}({\bf
C}^{\circ}(t)),I_{2}({\bf C}^{\circ}(t))\Bigr ).
\end{eqnarray*}
It follows from Eqs. (18), (23) and (24) that
\begin{equation}
\frac{dW}{dt}(t) = J(t)\frac{dT}{dt}(t)+{\bf A}(t)\cdot\frac{d{\bf
a}}{dt}(t) +2{\bf H}(t):{\bf D}(t)-Y(t),
\end{equation}
where
\[
{\bf H}(t) = (1-\mu_{\rm L}){\bf \Lambda}_{\rm B}^{\circ}(t)
+\mu_{\rm L}\Bigl \{ {\bf \Lambda}_{\rm
L}^{\circ}(t)\int_{0}^{\infty} \exp \Bigl [-\int_{0}^{t}
\Gamma(s,\omega)ds\Bigr ] p(\omega)d\omega +{\bf \Lambda}_{\rm
L}(t)\Bigr \}.
\]

\section{Constitutive equations}

We confine ourselves to quasi-static loadings with finite strains,
when mechanically induced changes in temperature, $\Delta T$, are
rather weak. This implies that the effect of temperature on the
specific heat, $c$, may be neglected. The following expression is
adopted for the free energy (per unit mass) of a filled elastomer:
\begin{equation}
\Psi=\Psi^{\circ}+(c-S^{\circ})(T-T^{\circ})-c T
\ln\frac{T}{T^{\circ}}+W,
\end{equation}
where $\Psi^{\circ}$ and $S^{\circ}$ are the free energy and the
entropy per unit mass in the stress-free state at the reference
temperature $T^{\circ}$, and the second and third terms on the
right-hand side of Eq. (26) describe the free energy of thermal
motion of chains.

For an incompressible medium, the Clausius--Duhem inequality reads \cite{CG67}
\begin{equation}
T \frac{dQ}{dt}=-S\frac{dT}{dt}-\frac{d\Psi}{dt}
+\frac{1}{\rho}\Bigl ( {\bf \Sigma}^{\prime}:{\bf D}-\frac{1}{T}{\bf q}\cdot
{\bf \nabla} T \Bigr) \geq 0,
\end{equation}
where $\rho$ is mass density, ${\bf \Sigma}^{\prime}$ is the
deviatoric component of the Cauchy stress tensor ${\bf \Sigma}$,
${\bf q}$ is the heat flux vector, $S$ is the entropy per unit
mass, and $Q$ is the entropy production. Substituting Eq. (26)
into Eq. (27) and using Eq. (25), we find that
\begin{eqnarray}
T \frac{dQ}{dt}&=& -\Bigl ( S-S^{\circ}-c\ln\frac{T}{T^{\circ}}+J\Bigr )\frac{dT}{dt}
+\frac{1}{\rho}\Bigl ( {\bf \Sigma}^{\prime}-2\rho {\bf H}\Bigr ):{\bf D}
\nonumber\\
&& +{\bf A}\cdot\frac{d{\bf a}}{dt}+Y-\frac{1}{\rho T}{\bf q}\cdot
{\bf \nabla} T \geq 0.
\end{eqnarray}
Because Eq. (28) is to be satisfied for an arbitrary loading
program, the expressions in brackets vanish. This implies the
formula for the entropy per unit mass
\[ S(t)=S^{\circ}+c\ln\frac{T(t)}{T^{\circ}}-J(t) \]
and the constitutive equation
\begin{eqnarray}
{\bf \Sigma}(t) &=& -P(t){\bf I}+2\rho \Bigr \{ (1-\mu_{\rm
L}){\bf \Lambda}_{\rm B}^{\circ}(t)
\nonumber\\
&&+\mu_{\rm L}\Bigl [ {\bf \Lambda}_{\rm L}^{\circ}(t)\int_{0}^{\infty}
\exp \Bigl (-\int_{0}^{t} \Gamma(s,\omega)ds\Bigr ) p(\omega)d\omega
+{\bf \Lambda}_{\rm L}(t) \Bigr ]\Bigr \},
\end{eqnarray}
where $P$ is pressure.

We adopt the Fourier law for the heat flux vector ${\bf q}$,
\[ {\bf q}=-\kappa {\bf \nabla}T \]
where $\kappa>0$ is thermal diffusivity. This implies that the
last term on the right-hand side of Eq. (28) in non-negative. It
follows from Eq. (19) that the functional $Y$ is non-negative as
well. To ensure that the Clausius--Duhem inequality (28) is
satisfied, it suffices to assume the derivative of ${\bf a}$ to be
proportional to the vector ${\bf A}$,
\[ \frac{d{\bf a}}{dt}=\eta {\bf A}, \]
where $\eta\geq 0$ is a function to be determined.

To distinguish explicitly rearrangement of temporary networks in
RLCJs and rupture of filler aggregates, we suppose that $\eta$ is
a function of the vector ${\bf a}$, the rate-of-strain tensor
${\bf D}$, and the Cauchy deformation tensor ${\bf C}^{\circ}$,
$\eta=\eta({\bf a},{\bf D}, {\bf C}^{\circ})$, which vanishes for
any time-independent straining of a specimen,
\[
\eta({\bf a},{\bf D}, {\bf C}^{\circ})\Bigl |_{{\bf D}={\bf
0}}=0.
\]
In the sequel, we concentrate on the study of tensile relaxation
tests and do not dwell on the kinetic equations for the vector of
internal variables.

\section{Uniaxial tension of a specimen}

We apply constitutive equation (29) to analyze stresses in a bar
under tension. Points of the bar refer to Cartesian coordinates
$\{ X_{i} \}$ in the stress-free state and to coordinates $\{
x_{i} \}$ in the deformed state, $(i=1,2,3)$. Uniaxial tension of
an incompressible medium is described by the formulas
\[
x_{1}=k(t)X_{1},
\qquad x_{2}=k^{-\frac{1}{2}}(t) X_{2}, \qquad
x_{3}=k^{-\frac{1}{2}}(t) X_{3},
\]
where $k=k(t)$ is the extension ratio. The Finger tensor, ${\bf
F}^{\circ}(t)$, and the relative Finger tensor, ${\bf F}(t,\tau)$,
read
\begin{eqnarray}
{\bf F}^{\circ}(t) &=& k^{2}(t){\bf e}_{1}{\bf e}_{1}+\frac{1}{k(t)}
({\bf e}_{2}{\bf e}_{2}+{\bf e}_{3}{\bf e}_{3}),
\nonumber\\
{\bf F}(t,\tau) &=& \biggl (\frac{k(t)}{k(\tau)}\biggr )^{2} {\bf
e}_{1}{\bf e}_{1}+\frac{k(\tau)}{k(t)} ({\bf e}_{2}{\bf
e}_{2}+{\bf e}_{3}{\bf e}_{3}),
\end{eqnarray}
where ${\bf e}_{i}$ are base vectors of the frame $\{ X_{i} \}$.
Substituting expressions (30) into Eq. (29), we find the non-zero
components $\Sigma_{k}$ ($k=1,2$) of the Cauchy stress tensor
\[ {\bf \Sigma}=\Sigma_{1}{\bf e}_{1}{\bf e}_{1}
+\Sigma_{2} ({\bf e}_{2}{\bf e}_{2}+{\bf e}_{3}{\bf e}_{3}). \]
Excluding the unknown pressure $P(t)$ from the expressions for
$\Sigma_{k}$ with the help of the boundary condition
\[ \Sigma_{2}=0, \]
we find the longitudinal stress
\begin{eqnarray}
\Sigma_{1}(t) &=& 2\rho \biggl \{ (1-\mu_{\rm L}) \phi_{\rm B\,
1}^{\circ}(t) +\mu_{\rm L}\phi_{\rm L\, 1}^{\circ}(t)
\int_{0}^{\infty} \exp \Bigl [-\int_{0}^{t}
\Gamma(s,\omega)ds\Bigr ] p(\omega)d\omega
\nonumber\\
&&+\Bigl [(1-\mu_{\rm L}) \phi_{\rm B\, 2}^{\circ}(t) +\mu_{\rm
L}\phi_{\rm L\, 2}^{\circ}(t) \int_{0}^{\infty} \exp \Bigl
(-\int_{0}^{t} \Gamma(s,\omega)ds\Bigr ) p(\omega)d\omega \Bigr ]
\nonumber\\
&&\times \Bigl [k^{2}(t)+k^{-1}(t)\Bigr ]
\biggr \} \Bigl [k^{2}(t)-k^{-1}(t)\Bigr ]
\nonumber\\
&& +2\rho \mu_{\rm L}\int_{0}^{\infty} p(\omega) d\omega \int_{0}^{t}
\biggl \{ \phi_{\rm L\, 1}(t,\tau)
+\phi_{\rm L\, 2}(t,\tau)\biggl [ \biggl (\frac{k(t)}{k(\tau)}\biggr )^{2}
+\frac{k(\tau)}{k(t)}\biggr ]\biggr \}
\nonumber\\
&&\times
\biggl [ \biggl (\frac{k(t)}{k(\tau)}\biggr )^{2}-\frac{k(\tau)}{k(t)}\biggr ]
\Gamma(\tau,\omega)\exp \Bigl [-\int_{\tau}^{t} \Gamma(s,\omega)ds\Bigr ]d\tau.
\end{eqnarray}
Equation (31) describes the stress in a specimen for an arbitrary
loading program. To compare results of numerical simulation with
experimental data, we focus on the relaxation test,
\begin{equation}
k(t)=\left \{ \begin{array}{lll}
1, && t<0,\\
\lambda, && t>0,
\end{array}
\right .
\end{equation}
where $\lambda >1$ is a constant. Substitution of Eq. (32) into
Eq. (31) implies that
\[ \Sigma_{1}(t)=\Bigl [ \sigma_{\rm B}+\sigma_{\rm L}-\sigma(t)\Bigr
] (\lambda^{2}-\lambda^{-1}), \]
where the quantities
\begin{eqnarray*}
\sigma_{\rm B} &=& 2\rho (1-\mu_{\rm L}) \Bigl [ \phi_{\rm
B\,1}^{\circ} + \phi_{\rm
B\,2}^{\circ}(\lambda^{2}+\lambda^{-1})\Bigr ],
\nonumber\\
\sigma_{\rm L} &=& 2\rho \mu_{\rm L} \Bigl [ \phi_{\rm L\,1}^{\circ}
+ \phi_{\rm L\,2}^{\circ}(\lambda^{2}+\lambda^{-1})\Bigr ],
\end{eqnarray*}
are independent of time and
\[
\sigma(t) = \sigma_{\rm L}\int_{0}^{\infty}\biggl [ 1
-\exp\biggl (-\int_{0}^{t} \Gamma(s,\omega)ds \biggr )\biggr ] p(\omega)d\omega.
\]
Bearing in mind Eq. (2), we present the dimensionless ratio
\[ R(t)=\frac{\Sigma_{1}(t)}{\Sigma_{1}(0)} \]
in the form
\begin{equation}
R(t)= 1-A\int_{0}^{\infty}\biggl [ 1 -\exp\biggl (-\Gamma_{\ast}
\exp(-\omega)t \biggr )\biggr ]p(\omega)d\omega,
\end{equation}
where
\begin{equation}
A=\frac{\sigma_{\rm L}}{\sigma_{\rm B}+\sigma_{\rm L}}.
\end{equation}
Equations (3) and (33) are determined by 4 adjustable parameters:
$A$ is the ratio of the relaxing stress to the total stress at the
beginning of the test, $\Gamma_{\ast}$ is the average rate of
relaxation, whereas $\Omega$ and $\Sigma$ in Eq. (3) for the
function $p(\omega)$ describe the distribution of relaxation
times. These constants are found by fitting experimental data in
relaxation tests with various elongation ratios, $\lambda$, for
unfilled and particle-reinforced rubbers.

\section{Experimental}

To assess the time-dependent response of particle-reinforced
elastomers, a series of uniaxial relaxation tests were carried out
at a constant temperature. Dumbbell specimens were provided by
TARRC laboratories (Hertford, UK) and were used as received
without mechanical or thermal pre-treatment prior to testing.

The first compound on the base of natural rubber (EDS--19)
contains 1 phr (parts per hundred parts of rubber) of high
abrasion furnace black (N--330) for UV protection. The filler
content is too low to affect the mechanical response
significantly, and we refer to this compound as an unfilled
rubber.

The other compound (EDS--16) contains 45 phr of carbon black (CB)
and this compound is treated as a filled rubber. Before the
vulcanization process, a sheet of rubber from which the specimens
were cut out was milled in a rolling mill to destroy agglomerates
of filler. This procedure leads to a weak anisotropy of the
material. The dumbbell specimens cut in the direction of milling
are referred to as R1, whereas the samples cut in the orthogonal
direction are abbreviated as R2.

A detailed formulation of compounds EDS--16 and EDS--19 is
presented in Table 1. The mean diameter of CB particles is
estimated as 30 nm, and the surface area of filler is about 80
m$^{2}$/g \cite{JM01}.

Relaxation tests were performed at room temperature at three
elongation ratios ($\lambda=1.2$, $\lambda=1.4$ and $\lambda=1.8$)
for samples made of the unfilled rubber, and at four elongation
ratios ($\lambda=1.2$, $\lambda=1.4$, $\lambda=1.8$ and
$\lambda=2.0$) for specimens made of the filled rubbers R1 and R2.
The experiments were carried out by using a testing machine
designed at the Institute of Physics (Vienna, Austria) and
equipped by a video-controlled system. To measure the longitudinal
strain, two reflection lines were drawn in the central part of
each specimen before loading (with the distance 7 mm between
them). Changes in the distance between these lines were measured
by a video-extensometer (with the accuracy of about 1 \%). The
tensile force was measured by using a standard loading cell. The
nominal stress was determined as the ratio of the axial force to
the cross-sectional area of a specimen in the stress-free state: 4
mm $\times$ 2 mm for the unfilled rubber and 4 mm $\times$ 1 mm
for the unfilled rubbers.

To analyze the effects of mechanical pre-loading and thermal
recovery, the following testing procedure was chosen:
\begin{itemize}
\item
Any specimen was loaded with the strain rate
$\dot{\epsilon}_{+}=5.0 \cdot 10^{-3}$ (s$^{-1}$) up to a given
elongation ratio $\lambda$, which was preserved constant during
the relaxation test (1 hour), and unloaded with the strain rate
$\dot{\epsilon}_{-}=-5.0 \cdot 10^{-3}$ (s$^{-1}$).

\item
The specimen was annealed for 24 hours.

\item
The specimen was loaded with the strain rate $\dot{\epsilon}_{+}$
up to the elongation ratio $\lambda_{\max}=3.0$ and unloaded with
the strain rate $\dot{\epsilon}_{-}$.

\item
The specimen was annealed for 24 hours.

\item
The specimen was loaded with the strain rate $\dot{\epsilon}_{+}$
up to the same elongation ratio, $\lambda$, as in the first test.
This elongation ratio was preserved constant during the relaxation
test (1 hour), and, afterwards, the specimen was unloaded with the
strain rate $\dot{\epsilon}_{-}$.

\item
The specimen was recovered in an oven for 24 hours at the constant
temperature $T=100$ $^{\circ}$C and cooled down by air to room
temperature for 4 hours.

\item
The specimen was loaded with the strain rate $\dot{\epsilon}_{+}$
up to the same elongation ratio as in the first test. This
elongation ratio was preserved constant during the relaxation test
(1 hour), and, afterwards, the specimen was unloaded with the
strain rate $\dot{\epsilon}_{-}$.
\end{itemize}

\section{Comparison with experimental data}

We begin with matching observations in the first relaxation test
on the unfilled rubber with the elongation ratio $\lambda=1.2$.
Because the parameters $\Gamma_{\ast}$ and $\Omega$ are mutually
dependent (an increase in $\Omega$ results in the growth of
$\Gamma_{\ast}$), we set $\Gamma_{\ast}=1.0$ (s$^{-1}$) in the
approximation of experimental data. Given a dimensionless ratio
$A$, the quantities $\Omega$ and $\Sigma$ are found by the
steepest-descent algorithm. The coefficient $A$ is determined by
the least-squares technique to ensure the minimum discrepancies
between experimental data and results of numerical simulation.

We proceed with matching observations in the first relaxation test
on the unfilled rubber with the elongation ratios $\lambda=1.4$
and $\lambda=1.8$ by using the parameters $\Omega$ and $\Sigma$
found in the approximation of experimental data at the elongation
ratio $\lambda=1.2$. The relaxation rate $\Gamma_{\ast}$ is
determined by the steepest-descent procedure, and the coefficient
$A$ is found by the least-squares method.

Figure 1 demonstrates fair agreement between observations and
results of numerical simulation. It is worth noting that the
relaxation curves at different elongation ratios, $\lambda$,
noticeably differ from each other, despite some scatter in the
experimental data.

Afterwards, we approximate experimental data in the second
relaxation test (after pre-loading) on the unfilled specimens with
the elongation ratios $\lambda=1.2$, $\lambda=1.4$ and
$\lambda=1.8$. In the fitting procedure, we employ the values of
$\Omega$ and $\Sigma$ which were obtained by matching observations
in the first relaxation test. The quantity $\Gamma_{\ast}$ is
determined by the steepest-descent algorithm, and $A$ is found by
the least-squares technique. Experimental data and results of
numerical analysis are plotted in Figure 2.

Finally, we fit experimental data in the third relaxation test
(after thermal recovery) on the unfilled rubber with the
elongation ratios $\lambda=1.2$, $\lambda=1.4$ and $\lambda=1.8$.
We use the same values of $\Omega$ and $\Sigma$ which were
employed in matching observations in the first and second
relaxation tests. The rate of breakage $\Gamma_{\ast}$ is found by
the steepest-descent method, and the dimensionless ratio $A$ is
determined by the least-squares technique. Figure 3 demonstrates
good agreement between experimental data and results of numerical
simulation.

Using the values of $A$ and $\Gamma_{\ast}$ found by fitting
observations, we calculate the average relaxation time,
\[ \tau_{\ast}=\Gamma_{\ast}^{-1}, \]
and the dimensionless parameter
\[ r=\frac{A}{1-A}. \]
It follows from Eq. (34) that $r$ equals the ratio of the stress
in regions with low concentrations of junctions, $\sigma_{\rm L}$,
to the stress in the bulk medium, $\sigma_{\rm B}$,
\[ r=\frac{\sigma_{\rm L}}{\sigma_{\rm B}}. \]
The relaxation time $\tau_{\ast}$ and the coefficient $r$ are
depicted in Figures 4 and 5 versus the first invariant of the
Cauchy deformation tensor, $I_{1}({\bf C}^{\circ})$. The effect of
straining on $\tau_{\ast}$ and $r$ is described by the linear
equations
\begin{equation}
\tau_{\ast}=\tau_{0}+\tau_{1}(I_{1}-3), \qquad
r=r_{0}+r_{1}(I_{1}-3),
\end{equation}
where the constants $\tau_{k}$ ($\tau_{0}\geq 0$) and $r_{k}$
($k=0,1$) are determined by the least-squares technique. Figures 4
and 5 show that phenomenological equations (35) provide an
acceptable accuracy of approximation.

We proceed with matching experimental data for the two grades (R1
and R2) of filled rubber by applying the same procedure of
fitting. Observations in relaxation tests and results of numerical
simulation are plotted in Figures 6 to 8 for the samples R1 and in
Figures 11 to 13 for the specimens R2. The parameters
$\tau_{\ast}$ and $r$ are depicted versus $I_{1}$ in Figures 9 and
10 for the rubber R1 and in Figures 14 and 15 for the rubber R2
together with the results of numerical analysis based on Eq. (35).

Adjustable parameters for various grades of unfilled and CB filled
rubbers are listed in Tables 2 to 4, where the abbreviations ``V",
``P" and ``T" are used to denote material parameters found in the
first (virgin specimens), second (pre-loaded specimens) and third
(specimens recovered at the elevated temperature) relaxation
tests, respectively.

\section{Discussion}

Figures 1 to 3, 6 to 8, and 10 to 12 reveal that the mechanical
responses of rubbers (both unfilled and particle-reinforced) in
the relaxation tests strongly depend on time. The relative
decrease in the longitudinal stress,
\[ \zeta(t_{0})=\frac{\Sigma_{1}(0)-\Sigma_{1}(t_{0})}{\Sigma_{1}(0)}, \]
after relaxation for $t_{0}=1$ hour at the minimum elongation
ratio $\lambda=1.2$ equals 6 \% for the unfilled rubber and varies
from 13 to 15 \% for the CB filled rubbers. The quantity
$\zeta(t_{0})$ for the filled rubbers (R1 and R2) noticeably
exceeds (about by twice) that for the unfilled rubber. The fact
that the presence of filler enhances relaxation of stresses in
elastomers may be explained by an increase in the number of
regions with low concentrations of junctions in a
particle-reinforced material compared to that in an unfilled
medium (loading of rubber with the filler particles results in an
extra level of inhomogeneity in the elastomeric matrix). However,
because the dimensionless ratios $\zeta(t_{0})$ have the same
order of magnitude for unfilled and filled rubbers, we suppose
that the viscoelastic behavior of elastomers cannot be described
exclusively by means of the mechanically-induced destruction of
filler clusters and slippage of macromolecules along the surfaces
of particles \cite{AH96,Han00}.

Fair agreement between experimental data in three relaxation tests
for the three grades of rubber and results of numerical simulation
with fixed values of $\Omega$ and $\Sigma$ leads us to the
conclusion that the relaxation spectrum (which is described in the
model by the distribution function, $p(\omega)$, for strands with
various energies of breakage, $\omega$) is not affected by
mechanical pre-loading and thermal treatment of specimens. The
adjustable parameters, $\Omega$ and $\Sigma$, accept similar
values for the two grades of filled rubber (R1 and R2), which
noticeably differ from those for the unfilled rubber, see Table 2.
At the same time, the width of the quasi-Gaussian distribution of
energies for breakage (which is characterized by the dimensionless
ratio $\xi$) is practically independent of the presence of filler
($\xi$ belongs to the interval between 0.5 and 0.6 for all
specimens). The average energy for breakage of strands, $\Omega$,
for the unfilled rubber exceeds by a factor of 2 (approximately)
that for the filled rubbers. This means that in the domains with
low concentrations of junctions where breakage and reformation of
strands occur, the strength of links between filler particles and
polymeric chains is essentially weaker than that for chemical and
physical cross-links and entanglements.

Figures 2, 6 and 9 demonstrate that the characteristic time of
relaxation, $\tau_{\ast}$, increases with the longitudinal
elongation $\lambda$. This conclusion contradicts the strain--time
superposition principle \cite{KE87,LK92}, which is conventionally
developed within the free-volume theory \cite{Doo51}. According to
that concept, mechanical loading of a specimen results in the
growth of the free volume, which, in turn, accelerates the
reformation process for mobile units (cooperatively rearranging
regions). The free-volume theory correctly describes the
time-dependent behavior of glassy polymers, where mechanically
induced changes in the specific volume are substantial, whereas
the applicability of this approach to rubbery polymers appears to
be questionable, because of the incompressibility of rubbers.

We assume that an increase in the relaxation time, $\tau_{\ast}$
(and the corresponding decrease in the relaxation rate,
$\Gamma_{\ast}$) with $\lambda$ may be associated with the
mechanically-induced crystallization of polymeric chains, see,
e.g., \cite{Dro99}.

In a stress-free state of a rubbery polymer, strands between
junctions are composed of a number of statistically independent
segments freely connected in sequel. For a flexible chain, the
total mechanical energy of these segments is negligible compared
to the configurational component of the free energy,
\[ \Delta \Psi_{\rm c}=-TS_{\rm c}. \]
Here $S_{\rm c}=k_{\rm B}\langle \ln N\rangle $ is the
configurational entropy per chain, $N$ is the number of
configurations available for a strand with a given end-to-end
vector, and the angle brackets stand for the mean value.
Stretching of a strand causes an increase in the absolute value of
the end-to-end vector, which results in a substantial decrease in
the number of configurations, $N$. As a consequence, $S_{\rm c}$
decreases and the configurational component of the free energy,
$\Delta \Psi_{\rm c}$, increases. At some elongation ratio,
$\lambda$, the free energy for a flexible strand exceeds that for
a partially crystallized one (where some segments are organized in
a crystallite structure, whereas the others remain freely
jointed), and the crystallization process begins. Because
stress-induced crystallization of macromolecules is a kinetic
process, it begins from some nuclei of crystallites (segments with
restricted mobility). The junctions between chains are supposed to
play the role of these nuclei, which implies that the growth of
crystallites occurs from the ends of a strand to its center.

Crystallization of strands in the neighborhoods of junctions
substantially increases their strength, which results in a
decrease in the rate of breakage, $\Gamma_{\ast}$ (and an increase
in the relaxation time, $\tau_{\ast}$) with the elongation ratio.

It is worth noting an important difference between the
crystallization processes in unfilled and particle-reinforced
rubbers.

In an unfilled elastomer, all junctions in RLCJs are formed by
chemical and physical cross-links and entanglements. It is assumed
that elongation of a strand between these junctions results in the
growth of crystallites, which practically disappear after
unloading and subsequent annealing at room temperature. After
stretching of the specimens to the elongation ratio $\lambda=3.0$,
the crystallization process during the second relaxation test
becomes substantially less pronounced, which is concluded from the
fact that the slope of curve 1 in Figure 4 exceeds that for the
curve 2 by a factor of 4. This ``training" effect (a decrease in
the degree of stress-induced crystallization after straining a
sample with a large amplitude) may be explained by
\begin{enumerate}
\item
mechanically-induced alignment of polymeric chains along the
direction of stretching, see, e.g., \cite{BE00},

\item onset of weak links between aligned macromolecules (based on van
der Waals forces between segments of different chains).
\end{enumerate}

We suppose that only a part of weak inter-chain links is
annihilated when the samples are annealed at room temperature,
which implies that the mobility of chains in pre-loaded specimens
(at the second relaxation test) is substantially lower than that
in virgin samples. As a consequence, the degree of stress-induced
crystallization of strands in RLCJs in the second relaxation test
is noticeably reduced compared to that in the first relaxation
test, which implies that the relaxation rate, $\Gamma_{\ast}$,
estimated at the second test exceeds that at the first test for
all elongation ratios.

Thermal recovery of a (pre-loaded) unfilled rubber at an elevated
temperature results in (partial) breakage of inter-chain links
driven by thermal fluctuations. Thermal treatment reduces the
level of alignment of previously stretched chains in RLCJs and
increases their ability to be crystallized. This conclusion is
confirmed by comparison of curves 2 and 3 in Figure 4. This figure
reveals that for all elongation ratios, $\lambda$, the relaxation
times, $\tau_{\ast}$, of the thermally treated specimens exceed
those for the untreated samples (which may serve as a
corroboration for our hypothesis, because the rate of breakage for
active strands is strongly affected by the degree of their
crystallinity).

For a particle-filled rubber, a part of junctions in a temporary
network is formed by chemical and physical cross-links, whereas
the other part is formed by isolated filler particles and small
aggregates of filler. We adopt a conventional concept for the
rubber--filler interaction, see, e.g., \cite{MA01}, according to
which filler particles in an elastomeric matrix are surrounded by
glassy-like cores which are formed at the stages of mixing and
vulcanization of rubbery compounds. These hard shells consist of
polymeric chains with highly restricted mobility rigidly connected
to particles of filler. By analogy with unfilled elastomers, we
suppose that strands bridged to the hard cores are partially
crystallized under stretching. It is assumed, however, that the
hard cores resist the crystallization process, which implies that
given an elongation ratio, the degree of crystallization in a
sample of a filled rubber is lower than that in an unfilled one.
On the other hand, at unloading of a specimen, the glassy-like
shells surrounding filler particles prevent annihilation of
micro-crystallites in active strands linked to these particles,
which implies that residual crystallites remain in a
particle-reinforced elastomer, in contrast to an unfilled rubber,
where crystallites totally disappear when a specimen returns to
its stress-free state.

The difference between unfilled and filled elastomers may be
explained within the concept of preferential segregation for
polymer chains in the vicinity of filler particles, see, e.g.,
\cite{CSJ01}. According to this approach, during the composite
processing, short chains are segregated to interfaces between the
elastomeric matrix and particles, whereas the concentration of
long chains in the host material is increased. This means that
short chains are mainly bridged to the hard cores. Because the
molecular mobility of these chains is reduced compared to the long
ones, the rate of growth for crystallites in the short strands is
essentially lower than that in the long strands, which is observed
in experiments as a very slow increase in the relaxation time,
$\tau_{\ast}$, with the elongation ratio, $\lambda$. Comparison of
Figure 4 with Figures 9 and 14 implies that the ``rate of growth"
for the relaxation time with the measure of deformation, $I_{1}-3$
[estimated by means of the parameter $\tau_{1}$ in Eq. (35)] for
the virgin specimens of the unfilled rubber exceeds that for the
virgin samples of the filled rubbers by a factor of 30.

Glassy-like shells surrounding filler particles are assumed to
prevent the crystallites formed in short strands under stretching
from their destruction, which implies that stretching a specimen
up to $\lambda=3.0$ and its subsequent unloading do not affect
noticeably the concentration of crystallites in
particle-reinforced rubbers. This is evidenced as a pronounced
increase in the relaxation times, $\tau_{\ast}$, measured in the
second relaxation tests on the specimens of filled rubbers
compared to that in the relaxation tests on the virgin samples,
see Figures 9 and 14.

Thermal recovery of a filled rubber at an elevated temperature
causes destruction of crystallites arisen under loading. This
process is supposed to be driven by (i) thermal fluctuations which
substantially accelerate breakage and reformation of chains linked
to the filler particles and (ii) thermal annihilation of
glassy-like cores around particles. As a result, the degree of
crystallinity in reinforced specimens is reduced after recovery,
and the relaxation rate, $\Gamma_{\ast}$, is increased. The degree
of crystallinity estimated in the third relaxation test exceeds,
however, that for virgin specimens (see curves 1 and 3 in Figures
9 and 14), which means that some residual crystallites remain even
after 24 hours of recovery at $T=100$ $^{\circ}$C.

Table 3 evidences that the parameter $\tau_{0}$ for the filled
rubber R2 (samples cut out in the direction orthogonal to the
direction of milling) noticeably exceeds that for the filled
rubber R1 (specimens cut in the direction of milling) in all three
relaxation tests. This may be explained by alignment of chains in
RLCJs during the milling procedure. Polymeric chains with the
end-to-end vectors directed along the direction of loading in the
relaxation tests are strongly aligned during milling, which
substantially reduces their mobility (because of the presence of
inter-chain links formed at alignment) and, as a consequence,
their ability to be crystallized. Because these chains bear the
main load at uniaxial extension of specimens, the average
relaxation rate for specimens R1 remains rather high. On the
contrary, alignment of polymeric chains with the end-to-end
vectors directed perpendicular to the direction of stretching for
specimens R2 unsubstantially affects the relaxation rate (these
chains are weakly strained in tensile relaxation tests, and their
mechanical response is negligible), whereas the chains directed
along the stretching direction are practically not aligned, and
the degree of their crystallization is similar to that for the
chains in a specimen not subjected to milling.

Figures 5, 10 and 15 demonstrate that for all grades of rubber the
dimensionless ratio $r$ increases with the longitudinal strain in
the first and third relaxation tests and decreases in the second
test. This means that in the first and third tests, the stress,
$\sigma_{\rm L}$, in RLCJs grows more rapidly (with the elongation
ratio $\lambda$) than the stress, $\sigma_{\rm B}$, in the bulk
material, while in the second test, the stress in the bulk medium
increases with $\lambda$ substantially more rapidly than the
stress, $\sigma_{\rm L}$, in domains with low concentrations of
junctions. This qualitative difference between the mechanical
response in the first and third tests, on the one hand, and in the
second test, on the other, is observed for both unfilled and
particle-reinforced rubbers. It may be associated with the
stress-induced alignment of chains.

Experimental data on virgin samples show that uniaxial tension of
rubber causes a more pronounced growth of the stress, $\sigma_{\rm
L}$, in RLCJs than the growth of the stress, $\sigma_{\rm B}$, in
the bulk material. This difference may be explained by the
mechanically-induced crystallization of active strands in the
regions with low concentrations of junctions (we presume that
stress-induced crystallization of chains either does not occur in
the bulk medium or takes place with a substantially lower rate
than in RLCJs).

After pre-loading to $\lambda=3.0$ and subsequent annealing at
room temperature, the stress in the bulk material grows with the
elongation ratio, $\lambda$, more rapidly than the stress in
RLCJs. This observation is associated with a higher level of
alignment of chains in the bulk material (which causes its
strain-hardening) than that in domains with low concentrations of
junctions. We assume that breakage and reformation of active
strands during annealing for 24 hours noticeably destroy alignment
of chains in RLCJs driven by the mechanical pre-loading, whereas
alignment of chains in the bulk material (where reformation of
strands does not take place) remains at the same level as
immediately after pre-loading.

Finally, after thermal recovery at an elevated temperature, the
stress in RLCJs increases with the elongation ratio more rapidly
that the stress in the bulk material, which is explained (as in
the case of virgin specimens) mechanically-induced crystallization
of strands in inclusions with low concentrations of junctions.

It is worth noting that the dimensionless ratio $r$ for the
unfilled rubber accepts the same values for the first and third
relaxation tests (because crystallites disappear in an unfilled
elastomer after unloading and thermal recovery), while the ratio
$r$ for the third test on filled rubbers exceeds that for the
first test (which confirms our assumption that some residual
crystallites remain in a particle-reinforced elastomer after
unloading).

Comparison of Figure 5, on the one hand, and Figures 10 and 15, on
the other, reveals that the dimensionless parameter $r$ for the
unfilled rubber exceeds (about by twice) that for the filled
rubbers. This fact may be explained by a higher level of
inhomogeneity of particle-reinforced elastomers compared to the
unfilled ones, which implies that RLCJs in filled rubbers are
noticeably ``looser" than those in unfilled rubbers with the same
polymeric matrix.

Fitting of experimental data implies practically the same values
of the dimensionless ratio $r$ for specimens R1 and R2, which
means that the effect of milling on this parameter is negligible.

\section{Concluding remarks}

Constitutive equations have been derived for the isothermal
viscoelastic behavior of filled elastomers at finite strains. With
reference to the concept of temporary networks, a filled elastomer
is treated as a network of long chains connected to junctions
(chemical and physical cross-links, entanglements and filler
particles). It is assumed that a particle-reinforced rubber is a
strongly non-homogeneous medium (with the characteristic length of
inhomogeneity of about 1 $\mu$m) which consists of an elastomeric
matrix with a normal concentration of cross-links, wherein
inclusions with low concentrations of junctions are distributed.
In the domains with low concentrations of cross-links, the average
number of junctions between polymeric chains (per unit mass) is
less than that in the bulk material because of local
inhomogeneities in the distribution of a cross-linker at mixing.

It is assumed that intermolecular forces in the bulk material
prevent slippage of chains from junctions (which are treated as
permanent), whereas in RLCJs active strands (whose ends are linked
to separate junctions) can slip from the temporary junctions as
these strands are thermally agitated. Breakage of an active strand
(its transition into a dangling one) occurs at a random instant
when one of its ends detaches from a junction. An active strand is
reformed at a random time when the free end of a dangling strand
captures some junction in its vicinity. The stress in a dangling
strand totally relax before this strand merges with the network,
which implies that the stress-free state of an active strand
coincides with the deformed state of the network at the instant of
their merging. Unlike conventional theories of transient networks,
we postulate that different strands in RLCJs have different
energies for breakage and describe these energies by the
quasi-Gaussian distribution function, $p(\omega)$, with two
adjustable parameters, $\Omega$ and $\Sigma$.

Constitutive equations for a filled rubber are developed by using
the laws of thermodynamics. These equations are applied to
determine stresses at uniaxial extension of a specimen. The
time-dependent response of a filled rubber in the standard
relaxation test is characterized by four adjustable parameters
which are found by fitting observations. To determine these
constants, a series of relaxation tests have been carried out on
specimens of unfilled and CB filled natural rubber (virgin
samples, samples pre-loaded up to the elongation ratio
$\lambda=3.0$, and specimens pre-loaded and recovered at an
elevated temperature). Fair agreement is demonstrated between
results of numerical simulation and observations in relaxation
tests (with the duration of $t_{0}=1$ hour) at various elongation
ratios, $\lambda$, in the region of moderate finite deformations,
$\lambda\in (1,2)$.

The following conclusions are drawn:
\begin{enumerate}
\item
The unfilled and particle reinforced rubbers demonstrate a
pronounced viscoelastic response. The relative decrease in
stresses, $\zeta(t_{0})$, for the filled rubbers exceeds that for
the unfilled rubber by twice. This difference may be ascribed to
an increase in the number of regions with low concentrations of
junctions caused by the presence of filler.

\item
The average energy for breakage of strands, $\Omega$, decreases
with an increase in the filler content, whereas the width of the
quasi-Gaussian distribution of energies, $\xi=\Sigma/\Omega$, is
independent of the content of carbon black. The distribution
function for the energies of breakage, $p(\omega)$, is not
affected by strains, mechanical pre-loading (up to the elongation
ratio $\lambda=3.0$) and thermal recovery.

\item
The characteristic time of relaxation, $\tau_{\ast}$, increases
with the elongation ratio, $\lambda$. The growth of $\tau_{\ast}$
with $\lambda$ is explained by the mechanically-induced
crystallization of natural rubber in the regions with low
concentrations of junctions.

\item
A qualitative difference is revealed in the influence of
mechanical pre-loading and thermal recovery on the relaxation
time, $\tau_{\ast}$, for unfilled and particle-reinforced
elastomers. The characteristic time of relaxation for the unfilled
rubber is substantially reduced after pre-loading, and it weakly
increases after thermal recovery at an elevated temperature. On
the contrary, mechanical pre-loading of the filled rubber causes a
noticeable growth of its relaxation time, which decreases after
thermal recovery. This difference is ascribed to the influence of
glassy-like cores around filler particles, which prevent
annihilation of crystallites in active strands bridged to the
particles of filler.
\end{enumerate}

\section*{Acknowledgements}

We would like to express our gratitude to Dr. K. Fuller (TARRC,
UK) for providing us with rubber specimens and to Mr. G. Seifritz
for his help in carrying out mechanical tests.

\newpage

\begin{center}
{\bf Table 1:} Chemical composition of rubbers (phr by weight)
\vspace*{6 mm}

\begin{tabular}{|l| r r |}\hline
Formulation & EDS--16 & EDS--19 \\ \hline\hline
Natural rubber         & 100 & 100\\
Zinc oxide             & 5   & 5  \\
Stearic acid           & 2   & 2  \\
Carbon black (N330)    & 45  & 1  \\
Process oil            & 4.5 & 0  \\
Antidegradant(HPPD)    & 3   & 3  \\
Antiozonant wax        & 2   & 2  \\
Accelerator (CBS)      & 0.6 & 0.6\\
Sulphur                & 2.5 & 2.5\\
\hline
\end{tabular}
\end{center}
\vspace*{3 mm}

\begin{center}
{\bf Table 2:} Adjustable parameters $\Omega$ and $\Sigma$
\vspace*{6 mm}

\begin{tabular}{|l| c c c|}\hline
Rubber & $\Omega$ & $\Sigma$ & $\xi$ \\
\hline\hline
Unfilled   & 13.80 & 7.00 & 0.51 \\
R1         & 5.30 &  2.80 & 0.53 \\
R2         & 5.13 &  3.18 & 0.62 \\
\hline
\end{tabular}
\end{center}
\vspace*{3 mm}

\begin{center}
{\bf Table 3:} Adjustable parameters $\tau_{0}$ and $\tau_{1}$

\vspace*{6 mm}

\begin{tabular}{|l| c c c | c c c|}\hline
Rubber & \multicolumn{3}{c|}{$\tau_{0}$} & \multicolumn{3}{c|}{$\tau_{1}$}
\\ \hline\hline
 & V & P & T & V & P & T \\ \hline
Unfilled   & 0.0    & 1.1866 & 1.9068 & 10.4615 & 2.5193    & 3.1182\\
R1         & 1.0114 & 1.7418 & 1.2543 & 0.1217 & 0.3390    & 0.1960\\
R2         & 1.3129 & 2.8106 & 1.9364 & 0.3073 & $-0.0226$ & 0.1598\\
\hline
\end{tabular}
\end{center}
\vspace*{3 mm}

\begin{center}
{\bf Table 4:} Adjustable parameters $r_{0}$ and $r_{1}$

\vspace*{6 mm}

\begin{tabular}{|l| c c c | c c c|}\hline
Rubber & \multicolumn{3}{c|}{$r_{0}$} & \multicolumn{3}{c|}{$r_{1}$}
\\ \hline\hline
 & V & P & T & V & P & T \\ \hline
Unfilled   & 0.3754 & 0.4674 & 0.3754 & 0.1964 & $-0.0354$ & 0.1964\\
R1         & 0.2008 & 0.2415 & 0.2647 & 0.0364 & $-0.0286$ & 0.0310\\
R2         & 0.2132 & 0.2655 & 0.2389 & 0.0416 & $-0.0453$ & 0.0495\\
\hline
\end{tabular}
\end{center}
\newpage

\subsection*{List of figures}

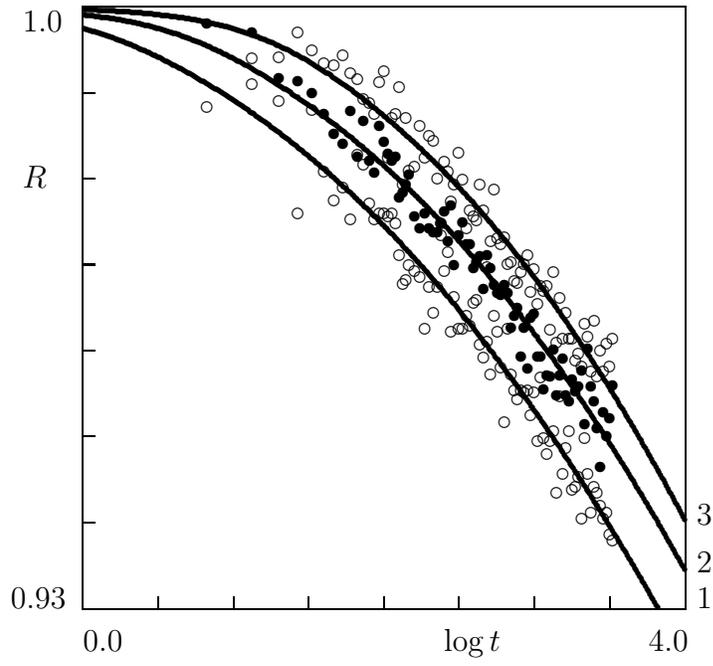
{\bf Figure 1:}
The dimensionless ratio $R$ versus time $t$ (s)
for unfilled rubber (virgin samples) in a relaxation test
with the elongation ratio $\lambda$.
Symbols: experimental data.
Solid lines: results of numerical simulation.
Curve 1: $\lambda=1.2$;
curve 2: $\lambda=1.4$;
curve 3: $\lambda=1.8$
\vspace*{2 mm}

\noindent
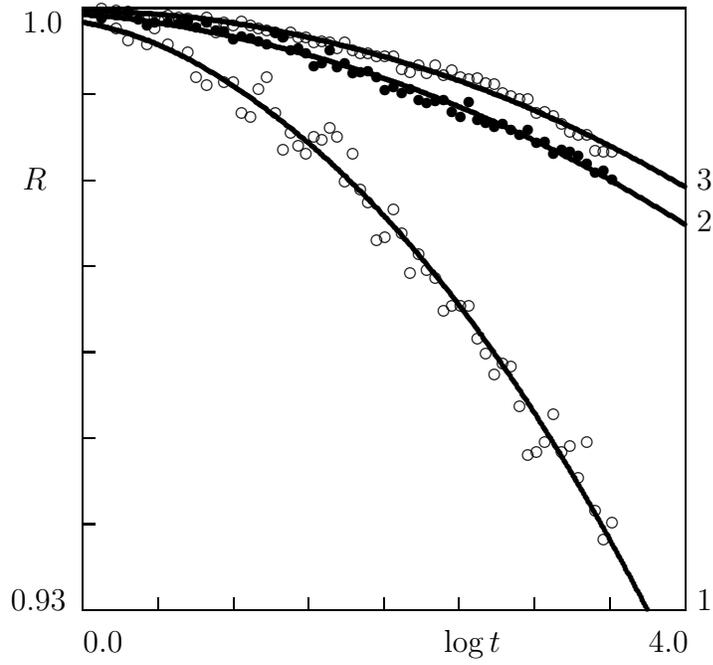
{\bf Figure 2:}
The dimensionless ratio $R$ versus time $t$ (s) for
unfilled rubber (after mechanical pre-loading) in a relaxation test
with the elongation ratio $\lambda$.
Symbols: experimental data.
Solid lines: results of numerical simulation.
Curve 1: $\lambda=1.2$;
curve 2: $\lambda=1.4$;
curve 3: $\lambda=1.8$
\vspace*{2 mm}

\noindent
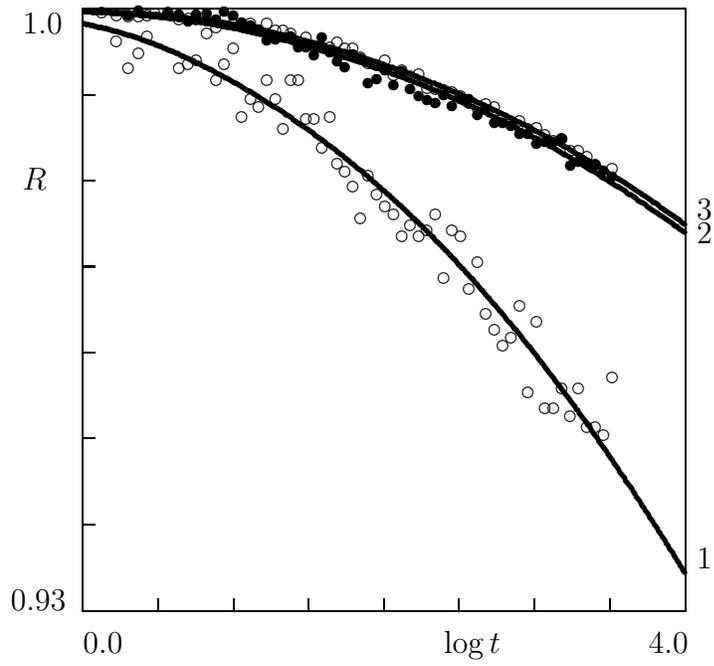
{\bf Figure 3:}
The dimensionless ratio $R$ versus time $t$ (s) for
unfilled rubber (after pre-loading and thermal recovery)
in a relaxation test with the elongation ratio $\lambda$.
Symbols: experimental data.
Solid lines: results of numerical simulation.
Curve 1: $\lambda=1.2$;
curve 2: $\lambda=1.4$;
curve 3: $\lambda=1.8$
\vspace*{2 mm}

\noindent
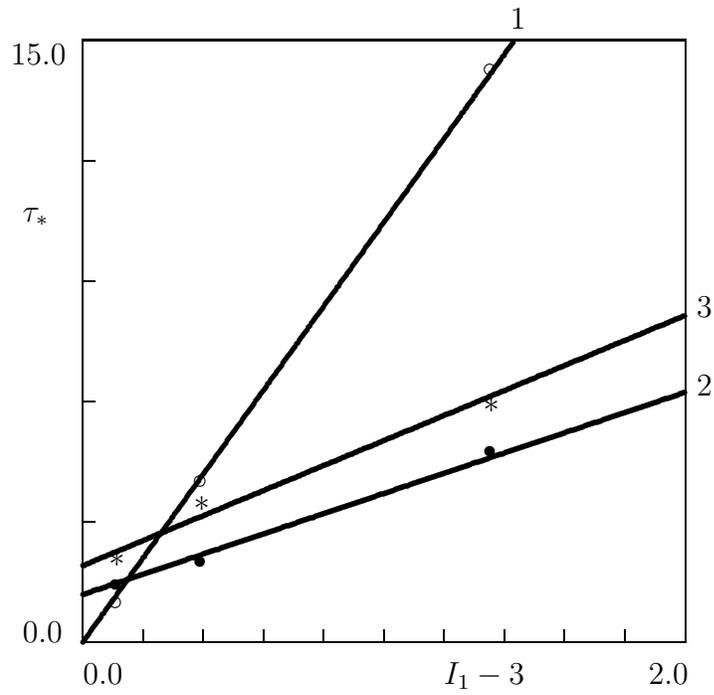
{\bf Figure 4:}
The characteristic time of relaxation $\tau_{\ast}$ (s)
versus the first invariant $I_{1}$ of the Cauchy deformation tensor
for unfilled rubber in relaxation tests.
Symbols: treatment of observations.
Unfilled circles: a virgin specimen;
filled circles: the specimen after mechanical pre-loading;
asterisks: the specimen after pre-loading and thermal recovery.
Solid lines: approximation of the experimental data by Eq. (35).
Curve 1: $\tau_{0}=0.0$, $\tau_{1}=10.4615$;
curve 2: $\tau_{0}=1.1866$, $\tau_{1}=2.5193$;
curve 3: $\tau_{0}=1.9068$, $\tau_{1}=3.1182$
\vspace*{2 mm}

\noindent
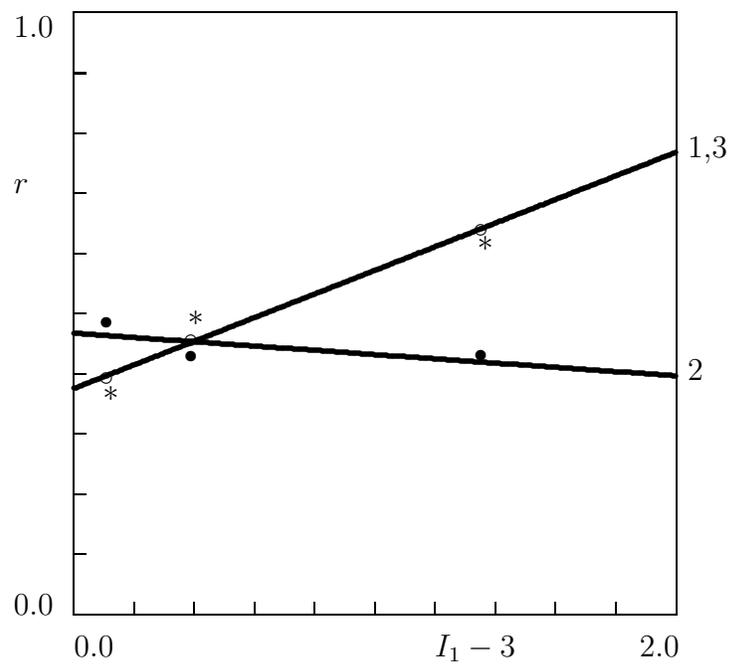
{\bf Figure 5:}
The dimensionless parameter $r$ versus the first invariant
of the Cauchy deformation tensor $I_{1}$ for unfilled rubber in relaxation tests.
Symbols: treatment of observations.
Unfilled circles: a virgin specimen;
filled circles: the specimen after mechanical pre-loading;
asterisks: the specimen after pre-loading and thermal recovery.
Solid lines: approximation of the experimental data by Eq. (35).
Curves 1 and 3: $r_{0}=0.3754$, $r_{1}=0.1964$;
curve 2: $r_{0}=0.4674$, $r_{1}=-0.0354$
\vspace*{2 mm}

\noindent
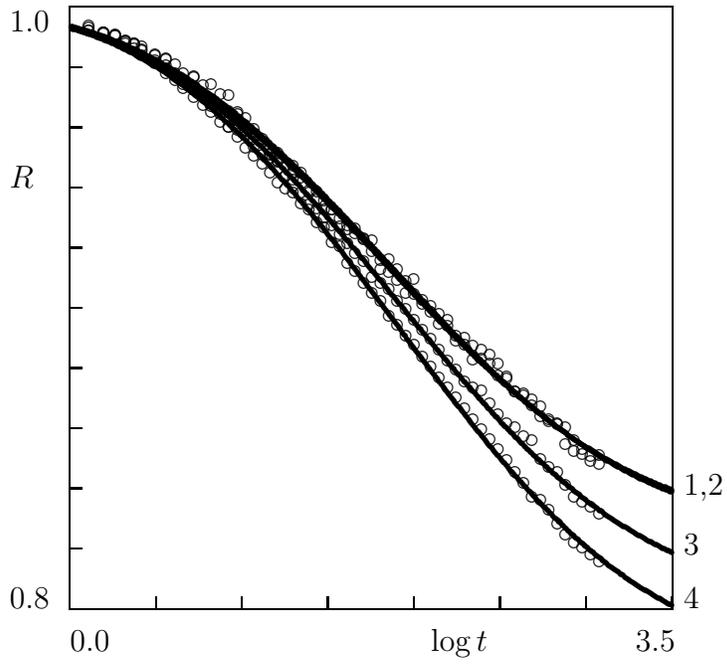
{\bf Figure 6:}
The dimensionless ratio $R$ versus time $t$ (s) for
CB filled rubber (R1) in a relaxation test with the elongation ratio $\lambda$.
Circles: experimental data.
Solid lines: results of numerical simulation.
Curve 1: $\lambda=1.2$;
curve 2: $\lambda=1.4$;
curve 3: $\lambda=1.8$;
curve 4: $\lambda=2.0$
\vspace*{2 mm}

\noindent
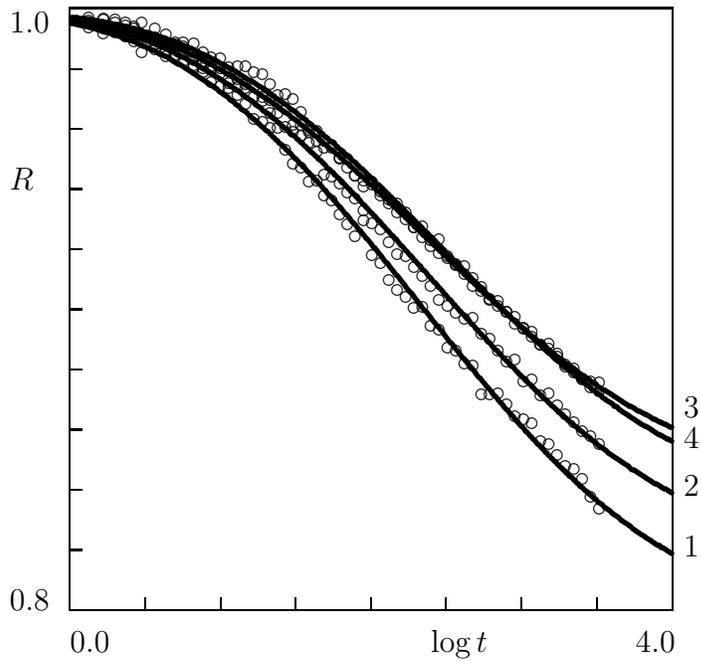
{\bf Figure 7:}
The dimensionless ratio $R$ versus time $t$ (s) for CB
filled rubber (R1) in a relaxation test (after mechanical pre-loading)
with the elongation ratio $\lambda$.
Circles: experimental data.
Solid lines: results of numerical simulation.
Curve 1: $\lambda=1.2$;
curve 2: $\lambda=1.4$;
curve 3: $\lambda=1.8$;
curve 4: $\lambda=2.0$
\vspace*{2 mm}

\noindent
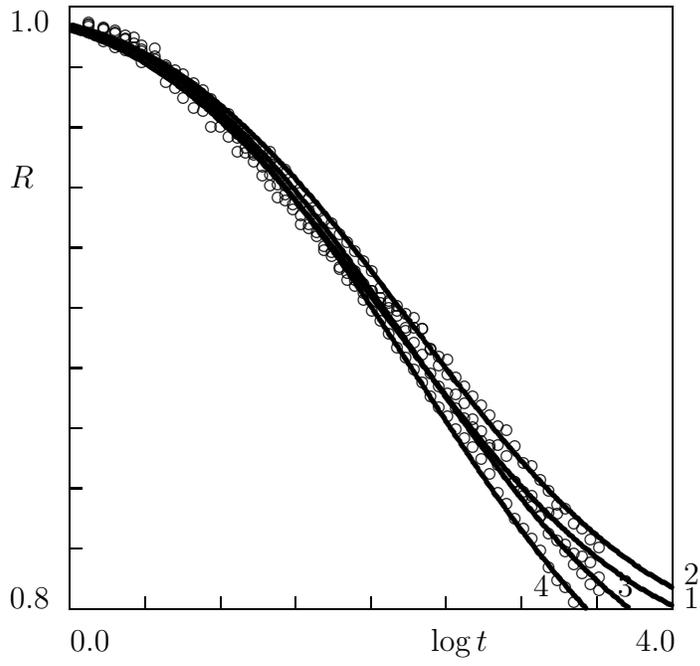
{\bf Figure 8:}
The dimensionless ratio $R$ versus time $t$ (s) for CB
filled rubber (R1) in a relaxation test (after pre-loading
and thermal recovery) with the elongation ratio $\lambda$.
Circles: experimental data.
Solid lines: results of numerical simulation.
Curve 1: $\lambda=1.2$;
curve 2: $\lambda=1.4$;
curve 3: $\lambda=1.8$;
curve 4: $\lambda=2.0$
\vspace*{2 mm}

\noindent
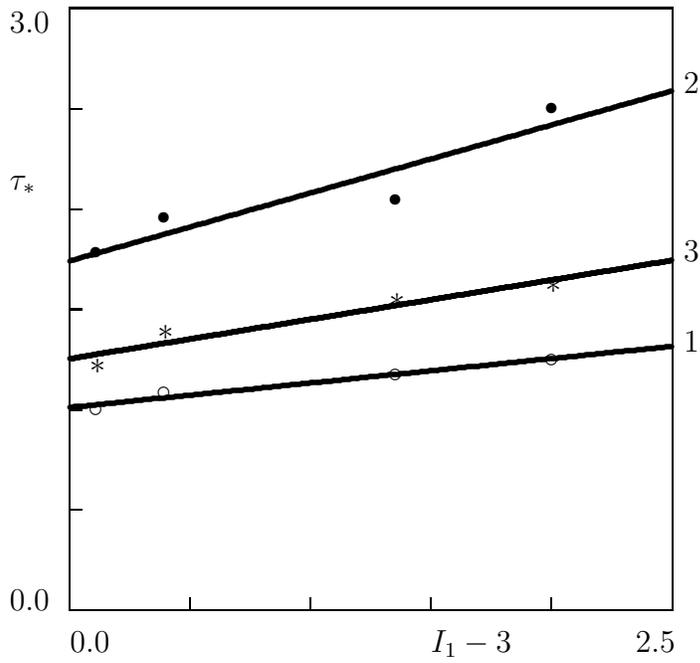
{\bf Figure 9:}
The characteristic time of relaxation $\tau_{\ast}$ (s) versus
the first invariant $I_{1}$ of the Cauchy deformation tensor for
CB filled rubber (R1) in relaxation tests.
Symbols: treatment of observations.
Unfilled circles: a virgin specimen;
filled circles: the specimen after mechanical pre-loading;
asterisks: the specimen after pre-loading and thermal recovery.
Solid lines: approximation of the experimental data by Eq. (35).
Curve 1: $\tau_{0}=1.0114$, $\tau_{1}=0.1217$;
curve 2: $\tau_{0}=1.7418$, $\tau_{1}=0.3390$;
curve 3: $\tau_{0}=1.2543$, $\tau_{1}=0.1960$
\vspace*{2 mm}

\noindent
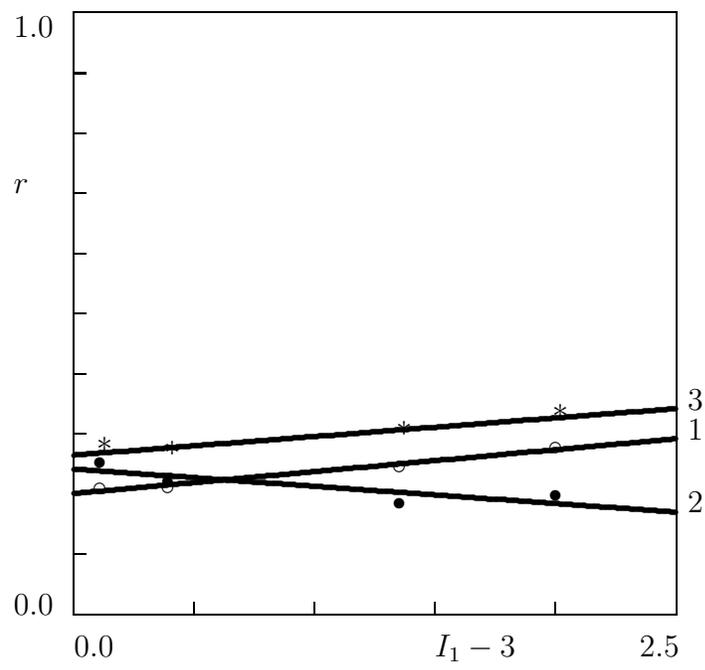
{\bf Figure 10:}
The dimensionless parameter $r$ versus the first
invariant of the Cauchy deformation tensor $I_{1}$ for
CB filled rubber (R1) in relaxation tests at room temperature.
Symbols: treatment of observations.
Unfilled circles: a virgin specimen;
filled circles: the specimen after mechanical pre-loading;
asterisks: the specimen after pre-loading and thermal recovery.
Solid lines: approximation of experimental data by Eq. (35).
Curve 1: $r_{0}=0.2008$, $r_{1}=0.0364$;
curve 2: $r_{0}=0.2415$, $r_{1}=-0.0286$;
curve 3: $r_{0}=0.2647$, $r_{1}=0.0310$
\vspace*{2 mm}

\noindent
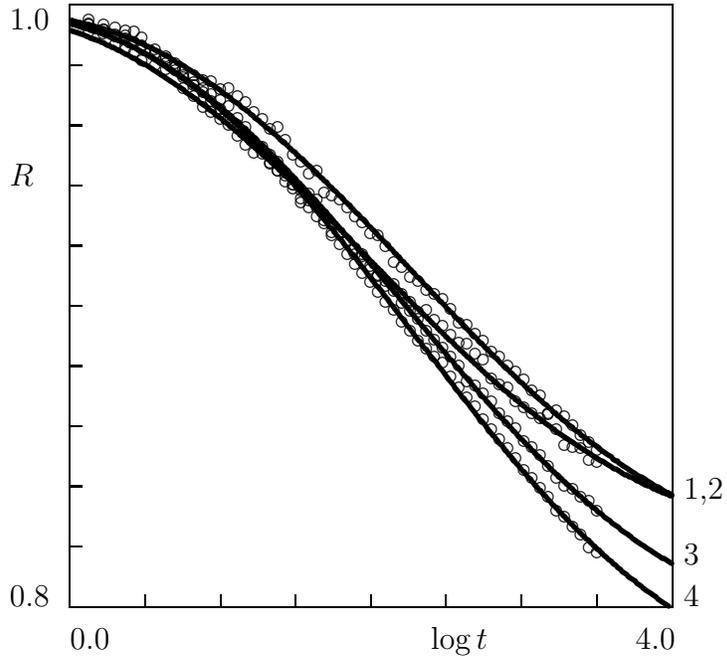
{\bf Figure 11:}
The dimensionless ratio $R$ versus time $t$ (s)
for CB filled rubber (R2) in a relaxation test
with the elongation ratio $\lambda$.
Circles: experimental data.
Solid lines: results of numerical simulation.
Curve 1: $\lambda=1.2$;
curve 2: $\lambda=1.4$;
curve 3: $\lambda=1.8$;
curve 4: $\lambda=2.0$
\vspace*{2 mm}

\noindent
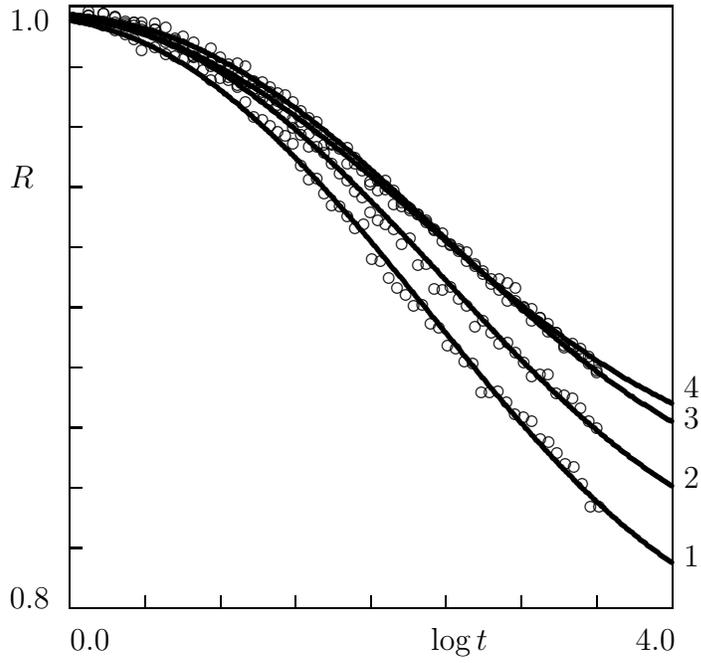
{\bf Figure 12:}
The dimensionless ratio $R$ versus time $t$ (s) for CB
filled rubber (R2) in a relaxation test (after mechanical pre-loading)
with the elongation ratio $\lambda$.
Circles: experimental data.
Solid lines: results of numerical simulation.
Curve 1: $\lambda=1.2$;
curve 2: $\lambda=1.4$;
curve 3: $\lambda=1.8$;
curve 4: $\lambda=2.0$
\vspace*{2 mm}

\noindent
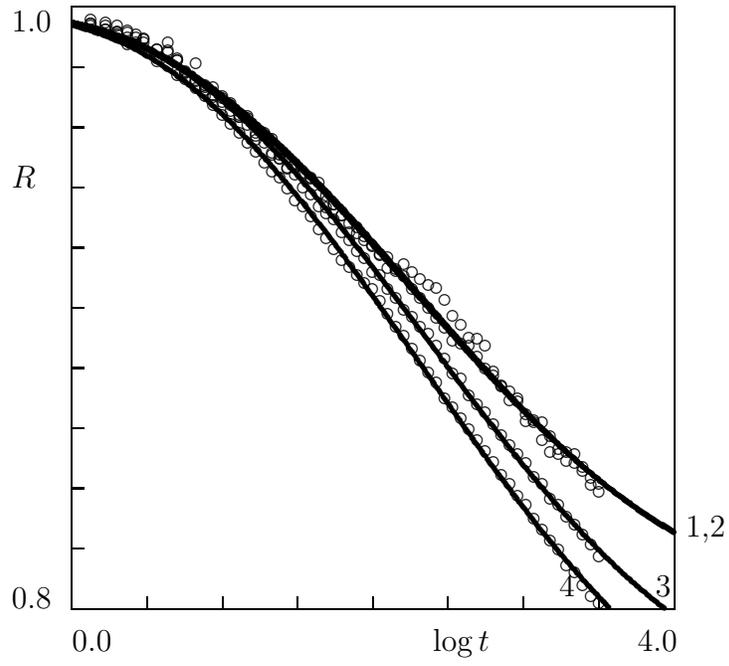
{\bf Figure 13:}
The dimensionless ratio $R$ versus time $t$ (s) for CB
filled rubber (R2) in a relaxation test (after pre-loading
and thermal recovery) with the elongation ratio $\lambda$.
Circles: experimental data.
Solid lines: results of numerical simulation.
Curve 1: $\lambda=1.2$;
curve 2: $\lambda=1.4$;
curve 3: $\lambda=1.8$;
curve 4: $\lambda=2.0$
\vspace*{2 mm}

\noindent
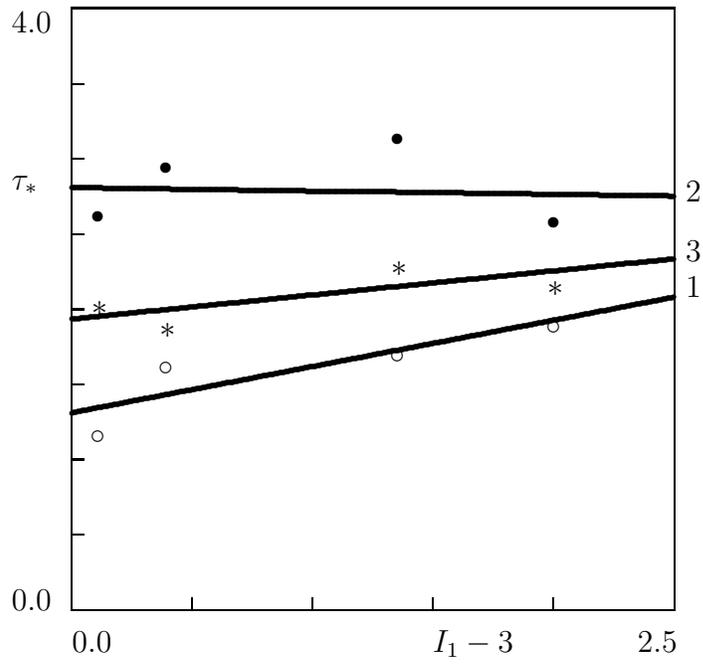
{\bf Figure 14:}
The characteristic time of relaxation $\tau_{\ast}$ (s) versus
the first invariant $I_{1}$ of the Cauchy deformation tensor for
CB filled rubber (R2) in relaxation tests.
Symbols: treatment of observations.
Unfilled circles: a virgin specimen;
filled circles: the specimen after mechanical pre-loading;
asterisks: the specimen after pre-loading and thermal recovery.
Solid lines: approximation of the experimental data by Eq. (35).
Curve 1: $\tau_{0}=1.3129$, $\tau_{1}=0.3073$;
curve 2: $\tau_{0}=2.8106$, $\tau_{1}=-0.0226$;
curve 3: $\tau_{0}=1.9364$, $\tau_{1}=0.1598$
\vspace*{2 mm}

\noindent
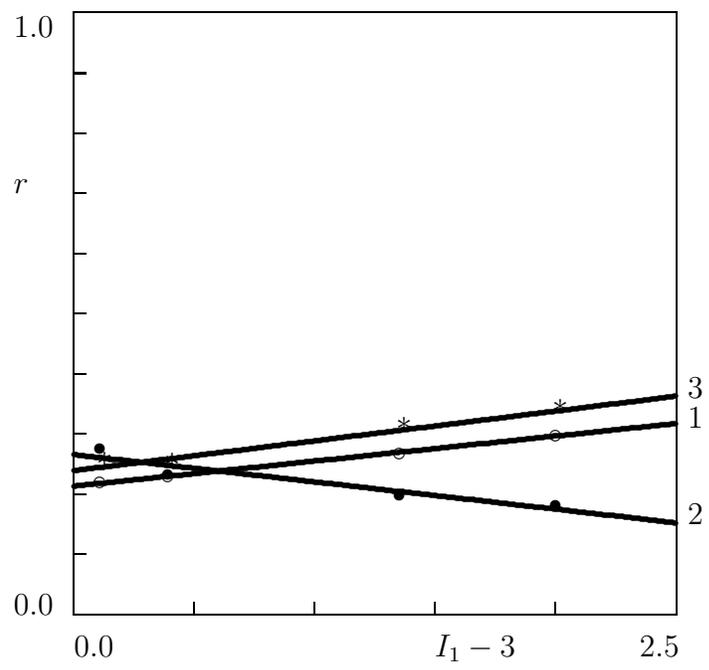
{\bf Figure 15:}
The dimensionless parameter $r$ versus the first
invariant of the Cauchy deformation tensor $I_{1}$ for
CB filled rubber (R2) in relaxation tests at room temperature.
Symbols: treatment of observations.
Unfilled circles: a virgin specimen;
filled circles: the specimen after mechanical pre-loading;
asterisks: the specimen after pre-loading and thermal recovery.
Solid lines: approximation of the experimental data by Eq. (35).
Curve 1: $r_{0}=0.2132$, $r_{1}=0.0416$;
curve 2: $r_{0}=0.2655$, $r_{1}=-0.0453$;
curve 3: $r_{0}=0.2389$, $r_{1}=0.0495$
\end{document}